\documentclass[aps,prd,twocolumns,eqsecnum,preprintnumbers,showpacs,amsmath,amssymb]
{revtex4}

\usepackage{graphicx}% Include figure files
\usepackage{dcolumn}% Align table columns on decimal point
\usepackage{bm}% bold math

\newcommand{\dd}{{\textrm d}}
\newcommand{\GeV}{{\textrm{GeV}}}
\newcommand{\MeV}{{\textrm{MeV}}}

\begin{document}

\title{ANALYTICAL DERIVATION AND NUMERICAL CALCULATION \\ OF THE $\alpha_s$-ORDER CORRECTION TO THE NON-PERTURBATIVE \\ EQUATION OF STATE FOR $SU(3)$ GLUON MATTER }

\vspace{3mm}

\author{V. Gogokhia$^{1,2}$, \ M. Vas\'uth$^1$}
\email[]{gogohia.vahtang@wigner.mta.hu}
\email[]{gogokhia@rmi.ge}
\email[]{vasuth.matyas@wigner.mta.hu}

\vspace{3mm}

\affiliation{$^1$WIGNER RCP, RMKI, Depart. Theor. Phys., Budapest 1121,
P.O.B. 49, H-1525, Hungary}

\affiliation{$^2$A. Razmadze Mathematical Inst. of I. Javakhishvili Tbilisi State University,
Depart. Theor. Phys., 2 University st., 0186 Tbilisi, Georgia}

\date{\today}
\begin{abstract}
In our previous works the effective potential approach for composite operators has been
generalized to non-zero temperature in order to derive the analytical equation
of state for pure $SU(3)$ Yang-Mills fields without quark degrees of freedom. In the absence of
external sources this is nothing but the vacuum energy density. The key element of this derivation is
the introduction of a temperature dependence into the expression for the bag constant. The non-perturbative
part of the analytical equation of state does not depend on the coupling constant, but instead introduces
a dependence on the mass gap. This is responsible for the large-scale dynamical structure of the QCD ground state.
The perturbative part of the analytical equation of state does depend on the QCD fine-structure coupling constant
$\alpha_s$. Here we develop the analytical formalism, incorporating the perturbative part in a self-consistent way. It makes it possible to calculate the PT contributions to the equation of state in terms of the convergent series in integer powers of a small $\alpha_s$. We also explicitly derive and numerically calculate
the first perturbative contribution of the $\alpha_s$-order to the non-perturbative part of the equation of state derived and calculated previously. The analytic equation of state or, equivalently, the gluon pressure is exponentially suppressed at low temperatures, while at the temperature $T=T_c = 266.5 \ \MeV$ it has a maximum, if
divided by $T^4/3$. It demonstrates a highly non-trivial dependence on the mass gap and the temperature
near $T_c$ and up to approximately $(3-4)T_c$. At very high temperatures its polynomial character is confirmed, containing the terms proportional to $T^2$ and $T$ with a non-analytical dependence on the mass gap.
\end{abstract}


%\keywords{}

\maketitle

\section{Introduction}

Up to now, lattice QCD remained the only practical method to investigate
QCD at finite temperature and baryon density \cite{1,2,3,4,5}. Recently it underwent a rapid
progress (\cite{6,7,8,9,10} and references therein). However, lattice QCD is primarily aimed
at obtaining well-defined corresponding expressions in order
to get realistic numbers for physical quantities. One may therefore get numbers and curves
without understanding what is the physics behind them. Such an understanding can
only come from the dynamical theory, which is continuous QCD. For
example, any description of the quark-gluon plasma (QGP) \cite{4,5} has to be formulated within the
framework of a dynamical theory. The need for an analytical equation of state (EoS)
remains, but of course it should be essentially non-perturbative
(NP), approaching the so-called Stefan-Boltzmann (SB) limit at
very high temperatures \cite{11,12}. Thus the approaches of analytic NP QCD and lattice QCD
to finite-temperature QCD do not exclude each other; on the contrary, they should be complementary.
This is especially true at low temperatures where the thermal QCD lattice calculations suffer from
big uncertainties (see above-cited papers), while any analytic NP approach has to correctly
reproduce thermal QCD lattice results at high temperatures.

The formalism we use to generalize it to non-zero temperature is the effective potential approach for composite operators \cite{13}. In the absence of external sources it is nothing but the vacuum energy density (VED). This approach is NP from the very beginning, since it deals with the expansion of the corresponding skeleton vacuum loop diagrams, and thus allows one to calculate the VED from first principles.
The key element in this program was the extension of our paper \cite{14} to non-zero temperature \cite{15}. This makes it possible to introduce the temperature-dependent bag constant (pressure) as a function of the mass gap which is responsible for the large-scale structure of the QCD ground state. The confining dynamics in the gluon matter (GM) will therefore be nontrivially taken into account directly through the mass gap and via the temperature-dependent bag constant itself, but other NP effects will also be present. Being NP, the effective approach for composite operators, nevertheless, makes it possible to incorporate the thermal perturbation theory (PT) expansion in a self-consistent way.

The main purpose of this article is the derivation of the analytical EoS for the GM, i.e., a system consisting purely of $SU(3)$ Yang-Mills (YM) fields without quark degrees of freedom.
Its NP part which solely depends on the mass gap has been evaluated in \cite{15}. Its PT part together with the mass gap depends on the QCD fine-structure constant $\alpha_s$.
Here we are going to formulate and develop the analytic formalism which makes it possible to determine the PT part of the YM EoS or, equivalently, the GM EoS in terms of the convergent series in integer powers of $\alpha_s$.
So this allows to calculate the PT contributions termwise in all orders of a small $\alpha_s$.
We also explicitly derive and numerically calculate the first PT contribution of the $\alpha_s$-order
to the NP part of the GM EoS derived and calculated previously \cite{15}. The low- and high-temperature expansions
for the GM EoS have analytically been evaluated.

The present paper is organized as follows. In section II the general expressions for the gluon pressure as a function of temperature are present. In section III all results for the NP part of the gluon pressure are collected and briefly explained. Thus in the short sections II and III we describe the results obtained previously in \cite{14,15}, and here we present them only for the reader's convenience in order to have a general picture in hand. In section IV the analytic thermal PT is developed, while in section
V we prove that it makes it possible to find the PT part of the gluon pressure in terms of the convergent series in integer powers of a small $\alpha_s$. The gluon pressure up to the $\alpha_s$-order term is derived in section VI.
In section VII we present our numerical results, where our conclusions are also given. And finally, in appendices A and B the low- and high-temperature expansions for the gluon pressure are analytically evaluated, respectively.

\section{The gluon pressure at non-zero temperature}

In the imaginary-time formalism \cite{12,16,17} all the
four-dimensional integrals can be easily generalized to non-zero
temperature $T$ according to the prescription (let us remind that in
the present investigation the signature is Euclidean from the very beginning)

\begin{equation}
\int {\dd q_0 \over (2\pi)} \rightarrow T \sum_{n=- \infty}^{+
\infty}, \quad \ q^2 = {\bf q}^2 + q^2_0 = {\bf q}^2 +
\omega^2_n = \omega^2 + \omega^2_n, \ \omega_n = 2n \pi T.
\end{equation}
In other words, each integral over $q_0$ of a loop momentum is to be
replaced by the sum over the Matsubara frequencies labeled by
$n$, which obviously assumes the replacement $q_0 \rightarrow
\omega_n= 2n \pi T$ for bosons (gluons).

Introducing the temperature dependence into the gluon pressure \cite{15}, we obtain

\begin{equation}
P_g(T) = P_{NP}(T) + P_{PT}(T) = B_{YM}(T) + P_{YM}(T) + P_{PT}(T),
\end{equation}
where the corresponding terms in frequency-momentum space are:

\begin{equation}
B_{YM}(T) = { 8 \over \pi^2}  \int_0^{\omega_{eff}} \dd\omega \ \omega^2 \ T
\sum_{n= - \infty}^{+ \infty} \left[ \ln \left( 1 + 3 \alpha^{INP}(\omega^2, \omega^2_n) \right) - {3 \over 4}
\alpha^{INP}(\omega^2, \omega^2_n) \right],
\end{equation}

\begin{equation}
P_{YM}(T) = - { 8 \over \pi^2}  \int_0^{\infty} \dd\omega \ \omega^2 \ T
\sum_{n= - \infty}^{+ \infty} \left[ \ln \left( 1 + {3 \over 4} \alpha^{INP}(\omega^2, \omega^2_n) \right) -
{3 \over 4} \alpha^{INP}(\omega^2, \omega^2_n) \right],
\end{equation}

\begin{equation}
P_{PT}(T) = - { 8 \over \pi^2}  \int_{\Lambda_{YM}}^{\infty} \dd\omega \ \omega^2 \ T
\sum_{n= - \infty}^{+ \infty} \left[ \ln \left( 1 + {3
\alpha^{PT}(\omega^2, \omega^2_n) \over 4 + 3
\alpha^{INP}(\omega^2, \omega^2_n)} \right) - {3 \over 4}
\alpha^{PT}(\omega^2, \omega^2_n) \right].
\end{equation}

In frequency-momentum space the intrinsically non-perturbative (INP) and the PT effective charges become

\begin{equation}
\alpha^{INP}(q^2) = \alpha^{INP}({\bf q}^2, \omega_n^2) =
\alpha^{INP}( \omega^2, \omega_n^2) = { \Delta^2 \over \omega^2 + \omega_n^2},
\end{equation}
and

\begin{equation}
\alpha^{PT}(q^2) = \alpha^{PT}({\bf q}^2, \omega_n^2) =
\alpha^{PT} (\omega^2, \omega_n^2) = { \alpha_s \over 1 + \alpha_s b
\ln (( \omega^2 + \omega_n^2) / \Lambda^2_{YM})},
\end{equation}
respectively. It is also convenient to introduce the following notations:

\begin{equation}
T^{-1} = \beta, \quad \omega = \sqrt{{\bf q}^2},
\end{equation}
where, evidently, in all the expressions ${\bf q}^2$ is the square of
the three-dimensional loop momentum, in complete agreement with the relations (2.1).

In Eq.~(2.6) $\Delta^2$ is the Jaffe-Witten mass gap \cite{18}, mentioned above, which is responsible for the large-scale structure of the QCD vacuum, and thus for its INP dynamics. Recently we have shown that confining
effective charge (2.6), and hence its $\beta$-function, is a result
of the summation of the skeleton (i.e., NP) loop diagrams, contributing to the full gluon self-energy in the $q^2 \rightarrow 0$ regime. This summation has been performed within the corresponding equations of motion
\cite{19} (and references therein).
In more detail (including the interpretation of Eq.~(2.6) and the explanation of all notations) the derivation of the bag constant as a function of the mass gap and its generalization to non-zero temperature has been completed in \cite{14} and \cite{15}, respectively. Let us only note that we omit the subscript "s" in $\alpha^{INP}(q^2)$ which has been used in \cite{15}.

The PT effective charge $\alpha^{PT}(q^2)$ (2.7) is the generalization to non-zero temperature of the renormalization group equation solution, the so-called sum of the main PT logarithms \cite{19,20,21,22}.
Here $\Lambda^2_{YM} = 0.09 \ \GeV^2$ \cite{23} is the asymptotic scale parameter for $SU(3)$ YM fields,
and $b=(11 / 4 \pi)$ for these fields, while the strong fine-structure constant is
$\alpha_s \equiv \alpha_s(m_Z) = 0.1184$ \cite{24}. In Eq.~(2.7) $q^2$ cannot go below $\Lambda^2_{YM}$, i.e.,
$\Lambda^2_{YM} \leq q^2 \leq \infty$, which has already been
symbolically shown in Eq.~(2.5). In our previous works \cite{14,15} the expression (2.7) was denoted
as $\alpha^{AF}(q^2)$. However, here we prefer to denote it as in Eq.~(2.7), leaving the notation
$\alpha^{AF}(q^2)$ for the asymptotic freedom (AF) relation $\alpha^{AF}(q^2) = 1 / b \ln ( q^2 / \Lambda^2_{YM})$ itself \cite{19,20,21,22}. One can recover it from Eq.~(2.7) in the $q^2 \rightarrow \infty$ limit.

The NP pressure $P_{NP}(T) = B_{YM}(T)+ P_{YM}(T)$ and the PT pressure $P_{PT}(T)$, and hence the gluon pressure $P_g(T)$ (2.2), are normalized to zero when the interaction is formally switched off, i.e., letting
$\alpha_s = \Delta^2=0$. This means that the initial normalization condition of the free PT vacuum to zero \cite{13,14,15} holds at non-zero temperature as well.

\section{$P_{NP}(T)$ contribution}

One of the attractive features of the
effective charge (2.6) is that it allows an exact summation over the Matsubara
frequencies in the NP pressure $P_{NP}(T)$ given by the sum of the integrals (2.3) and (2.4). Collecting all
analytical results obtained in our previous work \cite{15}, we can write

\begin{equation}
P_{NP}(T) = B_{YM}(T)+ P_{YM}(T) = {6 \over \pi^2} \Delta^2 P_1 (T) + {16 \over \pi^2} T
[P_2(T) + P_3(T) - P_4(T)],
\end{equation}
and

\begin{equation}
P_1(T) = \int_{\omega_{eff}}^{\infty} \dd \omega {\omega \over e^{\beta\omega} -1},
\end{equation}
while

\begin{eqnarray}
P_2(T) &=& \int_{\omega_{eff}}^{\infty} \dd \omega \ \omega^2
\ln \left( 1- e^{-\beta\omega} \right), \nonumber\\
P_3(T)&=& \int_0^{\omega_{eff}} \dd \omega \ \omega^2 \ln
\left( 1 - e^{- \beta\omega'} \right), \nonumber\\
P_4(T) &=& \int_0^{\infty} \dd \omega \ \omega^2 \ln \left( 1 -
e^{- \beta \bar \omega} \right),
\end{eqnarray}
where $\omega_{eff}=1 \ \GeV$ and the mass gap $\Delta^2= 0.4564 \ \GeV^2$ for $SU(3)$ gauge theory are
fixed \cite{14,15}. Then $\omega'$ and $\bar \omega$ are given by the relations

\begin{equation}
\omega' = \sqrt{\omega^2 + 3 \Delta^2} = \sqrt{\omega^2 +
m'^2_{eff}}, \quad m'_{eff}= \sqrt{3} \Delta = 1.17 \ \GeV,
\end{equation}
and

\begin{equation}
\bar \omega = \sqrt{\omega^2 + {3 \over 4} \Delta^2} =
\sqrt{\omega^2 + \bar m^2_{eff}}, \quad \bar m_{eff}= {\sqrt{3}
\over 2} \Delta = 0.585 \ \GeV,
\end{equation}
respectively. $P_{NP}(T)$ is shown in Fig. 1. It is worth reminding that in the NP pressure (3.1) the bag pressure $B_{YM}(T)$ (2.3) is responsible for the formation of the massive gluonic excitations $\omega'$ (3.4), while the YM part $P_{YM}(T)$ (2.4) is responsible for the formation of the massive gluonic excitations $\bar \omega$ (3.5).

Concluding, let us note that the so-called gluon mean number \cite{12}

\begin{equation}
N_g \equiv N_g(\beta, \omega) = {1 \over e^{\beta\omega} -1}, \quad \beta = T^{-1},
\end{equation}
which appears in the integrals (3.2)-(3.3), describes the distribution and correlation of massless gluons in the GM.
Replacing $\omega$ by  $\bar \omega$ and $\omega'$ we can consider the corresponding gluon mean numbers as describing
the distribution and correlation of the corresponding massive gluonic excitations in the GM, see integrals $P_3(T)$
and $P_4(T)$ in Eq.~(3.3). They are of NP dynamical origin, since their masses are due to the mass gap $\Delta^2$. All different gluon mean numbers range continuously from zero to infinity. We have the two different massless excitations, propagating in accordance with the integral (3.2), conventionally denoted as $\omega_1$, and the first of the integrals (3.3), conventionally denoted as $\omega_2$.
However, they are not free, since in the PT $\Delta^2=0$ limit they vanish (the composition
$[P_2(T) + P_3(T) - P_4(T)]$ becomes zero in this case).
So the NP pressure describes the four different gluonic excitations. The gluon mean numbers are
closely related to the pressure; in particular, its exponential suppression in the $T \rightarrow 0$ limit and the polynomial structure in the $T \rightarrow \infty$ limit is determined by the corresponding asymptotics of the gluon mean numbers. For the explicit evaluation of the low- and high-temperature expansions for the NP pressure (3.1) see appendices A and B, respectively.

\section{Analytic thermal PT }

Let us begin here reminding that we were able to perform the summation over the Matsubara frequencies analytically (i.e., exactly) for the NP part of the gluon pressure (2.2). To do the same for its PT part (2.5) is a formidable task. The only way to evaluate it is the numerical summation over $n$ and the integration over $\omega$, which is beyond our possibilities at present (if it is possible at all).
Our primary goal in this article is threefold. Firstly, to develop the analytical formalism which makes it possible
to calculate $P_{PT}(T)$ (2.5) termwise in integer powers of a small $\alpha_s$. Secondly, to calculate explicitly the
PT contribution of the $\alpha_s$-order. Thirdly, to derive the low- and high-temperature expansions for the gluon pressure.

For the first goal, it is convenient to re-write the integral (2.5) as follows:

\begin{equation}
P_{PT}(T) = - { 8 \over \pi^2}  \int_{\Lambda_{YM}}^{\infty} \dd\omega \ \omega^2 \ T
\sum_{n= - \infty}^{+ \infty} \left[ \ln [ 1 + x(\omega^2, \omega^2_n)] - {3 \over 4}
\alpha^{PT}(\omega^2, \omega^2_n) \right],
\end{equation}
where

\begin{equation}
x(\omega^2, \omega^2_n) = {3 \alpha^{PT}(\omega^2, \omega^2_n) \over 4 + 3
\alpha^{INP}(\omega^2, \omega^2_n)} = {3 \over 4} {( \omega^2 + \omega^2_n) \over M (\bar \omega^2, \omega_n^2)}
{ \alpha_s \over 1 +  \alpha_s \ln z_n},
\end{equation}
with the help of the expressions (2.6) and (2.7). Here

\begin{equation}
M(\bar \omega^2, \omega^2_n) = \bar \omega^2 + \omega^2_n, \quad
\ln z_n \equiv \ln z (\omega^2,\omega^2_n) = b \ln ((\omega^2 + \omega^2_n)/ \Lambda^2_{YM}),
\end{equation}
and $\bar \omega^2$ is given in Eq.~(3.5). Let us also note that in these notations

\begin{equation}
\alpha^{PT}(\omega^2, \omega^2_n) \equiv \alpha (z_n) = { \alpha_s \over 1 +  \alpha_s \ln z_n}.
\end{equation}

There is an interesting observation concerning the argument $x(\omega^2, \omega^2_n)$
of the logarithm $\ln [ 1 + x(\omega^2, \omega^2_n)]$ in the integral (4.1). At its lower limit
$\omega = \Lambda_{YM}$ and $n=0$ the argument (4.2) numerically becomes

\begin{equation}
x(\Lambda^2_{YM}) = {3 \alpha_s \Lambda^2_{YM} \over 4 \Lambda^2_{YM} + 3 \Delta^2} = 0.0185,
\end{equation}
and the numerical values of $\Delta^2$, $\alpha_s$ and $\Lambda^2_{YM}$
given above have already been used. The argument of the logarithm is really small (it is an order of magnitude smaller than $\alpha_s$ itself), and it will become even smaller and smaller with $\omega$ and $n$ going to infinity. This means that the logarithm $\ln [ 1 + x(\omega^2, \omega^2_n)]$ in the integral (4.1) is legitimated to expand in powers of small
$x(\omega^2, \omega^2_n)$ at any $n$ and in the whole range of the integration over $\omega$, that's $\infty \geq \omega \geq \Lambda_{YM}$. Doing so, one obtains \cite{25}

\begin{equation}
\ln [1 + x(\omega^2, \omega^2_n)] = - \sum_{m=1}^{\infty} { (-1)^m \over m} x^m (\omega^2, \omega^2_n),
\quad x(\omega^2, \omega^2_n) \ll 1.
\end{equation}

Extracting the first term in the expansion (4.6), the integral (4.1) becomes

\begin{equation}
P_{PT}(T) = - { 8 \over \pi^2}  \int_{\Lambda_{YM}}^{\infty} \dd\omega \ \omega^2 \ T
\sum_{n= - \infty}^{+ \infty} \left[ [ x(\omega^2, \omega^2_n) - {3 \over 4} \alpha (z_n)]
- \sum_{m=2}^{\infty} { (-1)^m \over m} x^m (\omega^2, \omega^2_n) \right].
\end{equation}

From now on it is instructive to separate the two terms in the integral (4.7) as follows:

\begin{equation}
P_{PT}(T) = P_{PT}(\Delta^2; T) + P'_{PT}(T),
\end{equation}
where

\begin{eqnarray}
P_{PT}(\Delta^2; T) &=& - { 8 \over \pi^2}  \int_{\Lambda_{YM}}^{\infty} \dd\omega \ \omega^2 \ T
\sum_{n= - \infty}^{+ \infty} [ x(\omega^2, \omega^2_n) - (3 /4) \alpha (z_n)] \nonumber\\
&=& { 9 \over 2 \pi^2} \Delta^2 \int_{\Lambda_{YM}}^{\infty} \dd\omega \ \omega^2 \ T \sum_{n= - \infty}^{+ \infty}
\left[ { 1 \over M(\bar \omega^2, \omega^2_n)} {\alpha_s \over 1 +  \alpha_s \ln z_n} \right],
\end{eqnarray}
on account of the relations (4.2)-(4.4), and

\begin{equation}
P'_{PT}(T) = { 8 \over \pi^2}  \int_{\Lambda_{YM}}^{\infty} \dd\omega \ \omega^2 \ T
\sum_{n= - \infty}^{+ \infty} \left[
 \sum_{m=2}^{\infty} { (-1)^m \over m} x^m (\omega^2, \omega^2_n) \right], \quad x(\omega^2, \omega^2_n) \ll 1.
\end{equation}

\subsection{ $P_{PT}(\Delta^2; T)$ contribution}

The principal difference between these two terms (4.9) and (4.10) is that the former term vanishes in the formal PT
$\Delta^2 =0$ limit, while the latter one survives it. Let us consider the first term in more detail.
The function $(1+ \alpha_s \ln z_n)^{-1}$ in the integral (4.9) can formally be replaced by the expansion
in integer powers of $\alpha_s$, namely

\begin{equation}
(1 +  \alpha_s \ln z_n)^{-1}= \sum_{k=0}^{\infty} (-1)^k \alpha_s^k \ln^k z_n,
\end{equation}
which converges to the corresponding function in the whole range $\Lambda_{YM} \leq \omega \leq \infty$ and at
any $n$ for $|\alpha_s \ln z_n | < 1$. The integral (4.9) can equivalently be re-written as follows:

\begin{equation}
P_{PT}(\Delta^2; T) = { 9 \over 2 \pi^2} \Delta^2 \alpha_s \int_{\Lambda_{YM}}^{\infty} \dd\omega \ \omega^2 \ T \sum_{n= - \infty}^{+ \infty} \left[ { 1 \over M(\bar \omega^2, \omega^2_n)} \sum_{k=0}^{\infty} (-1)^k \alpha_s^k \ln^k z_n \right],
\end{equation}
which makes it possible to present it as a sum in integer powers of $\alpha_s$, namely

\begin{equation}
P_{PT}(\Delta^2; T) =  \sum_{k=1}^{\infty} \alpha_s^k P_k (\Delta^2; T),
\end{equation}
where

\begin{equation}
P_k(\Delta^2; T) = { 9 \over 2 \pi^2} \Delta^2 \int_{\Lambda_{YM}}^{\infty} \dd\omega \ \omega^2 \ T \sum_{n= - \infty}^{+ \infty} \left[ { 1 \over M(\bar \omega^2, \omega^2_n)} (-1)^{k-1} \ln^{k-1} z_n \right].
\end{equation}
For example, the first term $P_1(\Delta^2; T)$ explicitly looks like

\begin{equation}
P_1(\Delta^2; T) = { 9 \over 2 \pi^2} \Delta^2 \int_{\Lambda_{YM}}^{\infty} \dd\omega \ \omega^2 \ T \sum_{n= - \infty}^{+ \infty} { 1 \over M(\bar \omega^2, \omega^2_n)},
\end{equation}
where $M(\bar \omega^2, \omega^2_n) = \bar \omega^2 + \omega^2_n$, and $\bar \omega^2$ itself is given in the relation (3.5). In this integral the summation over the Matsubara frequencies can be performed analytically (i.e., exactly)
with the help of the expression explicitly given in \cite{15}. Omitting all the derivation and dropping the $\beta$-independent terms \cite{12}, one obtains

\begin{equation}
P_1(\Delta^2; T) = { 9 \over 2 \pi^2} \Delta^2 \int_{\Lambda_{YM}}^{\infty} \dd\omega \ \omega^2 \
{ 1 \over \bar \omega} { 1 \over e^{\beta \bar \omega}  - 1}.
\end{equation}

\subsection{ $P'_{PT}(T)$ contribution}

On account of the relations (4.2)-(4.4), the integral (4.10) becomes

\begin{equation}
P'_{PT}(T) = { 8 \over \pi^2}  \int_{\Lambda_{YM}}^{\infty} \dd\omega \ \omega^2 \ T
\sum_{n= - \infty}^{+ \infty} \left[ \sum_{m=2}^{\infty} b_m(\omega^2, \omega^2_n) {\alpha_s^m \over (1 +  \alpha_s \ln z_n)^m } \right],
\end{equation}
where

\begin{equation}
b_m(\omega^2, \omega^2_n) = - {(-1)^m 3^m \over m 4^m} { (\omega^2 + \omega^2_n)^m \over M^m (\bar \omega^2, \omega^2_n)}.
\end{equation}

In complete analogy with the expansion (4.11) one gets \cite{25}

\begin{equation}
(1 +  \alpha_s \ln z_n)^{-m}= \sum_{k=0}^{\infty}c_k(m) \alpha_s^k \ln^k z_n,
\end{equation}
where

\begin{equation}
c_0(m) = 1, \quad c_p(m) = {1 \over p } \sum_{k=1}^p (km - p +k)(-1)^k c_{p-k}, \quad p \geq 1.
\end{equation}
What has been said in connection with the expansion (4.11) is valid for the expansion (4.19) as well.
So on its account, the integral (4.17) can equivalently be re-written as follows:

\begin{equation}
P'_{PT}(T) = { 8 \over \pi^2}  \int_{\Lambda_{YM}}^{\infty} \dd\omega \ \omega^2 \ T
\sum_{n= - \infty}^{+ \infty} \left[ \sum_{m=2}^{\infty} b_m(\omega^2, \omega^2_n) \alpha_s^m
\sum_{k=0}^{\infty}c_k(m) \alpha_s^k \ln^k z_n \right].
\end{equation}

Let us consider the coefficients $b_m(\omega^2, \omega^2_n)$ (4.18) in more detail. Noting that

\begin{equation}
(\omega^2 + \omega^2_n) = M (\bar \omega^2, \omega^2_n) - {3 \over 4} \Delta^2,
\end{equation}
these coefficients can be present as follows:.

\begin{eqnarray}
b_m(\omega^2, \omega^2_n) &=& - \left( -{3 \over 4} \right)^m { 1 \over m} { (\omega^2 + \omega^2_n)^m \over M^m (\bar \omega^2, \omega^2_n)} \nonumber\\
&=&- \left( -{3 \over 4} \right)^m { 1 \over m} \sum_{p=0}^m {m \choose p} M^{p-m}(\bar \omega^2, \omega^2_n) \left( - {3 \over 4} \Delta^2 \right)^{m-p} \nonumber\\
&=& - \left( -{3 \over 4} \right)^m { 1 \over m} \left[ 1 + \sum_{p=0}^{m-1} {m \choose p} M^{p-m}(\bar \omega^2, \omega^2_n) \left( - {3 \over 4} \Delta^2 \right)^{m-p} \right], \nonumber\\
\end{eqnarray}
and

\begin{equation}
{m \choose p} = {m(m-1)...(m-p+1) \over p!}, \quad {m \choose 0} = {m \choose m} =1
\end{equation}
are the binomials coefficients. Substituting the expression (4.23) into Eq.~(4.21), it becomes

\begin{equation}
P'_{PT}(T) = \tilde{P}_{PT}(T)+ P'_{PT}(\Delta^2;T),
\end{equation}
where

\begin{equation}
\tilde{P}_{PT}(T) = - { 8 \over \pi^2}  \int_{\Lambda_{YM}}^{\infty} \dd\omega \ \omega^2 \ T
\sum_{n= - \infty}^{+ \infty} \left[ \sum_{m=2}^{\infty} \left( -{3 \over 4} \right)^m {\alpha_s^m \over m}
\sum_{k=0}^{\infty}c_k(m) \alpha_s^k \ln^k z_n \right],
\end{equation}
and

\begin{equation}
P'_{PT}(\Delta^2;T) = -{ 8 \over \pi^2}  \int_{\Lambda_{YM}}^{\infty} \dd\omega \ \omega^2 \ T
\sum_{n= - \infty}^{+ \infty} \left[ \sum_{m=2}^{\infty} \left( -{3 \over 4} \right)^m {\alpha_s^m \over m} P^{(n)}_m(\Delta^2) \sum_{k=0}^{\infty}c_k(m) \alpha_s^k \ln^k z_n \right]
\end{equation}
with

\begin{equation}
P^{(n)}_m(\Delta^2) =
\sum_{p=0}^{m-1} {m \choose p} M^{p-m}(\bar \omega^2, \omega^2_n) \left( - {3 \over 4} \Delta^2 \right)^{m-p}.
\end{equation}

It is convenient to present the integral (4.26) in the following way

\begin{equation}
\tilde{P}_{PT}(T) = - { 9 \over 2 \pi^2} \alpha_s^2 \int_{\Lambda_{YM}}^{\infty} \dd\omega \ \omega^2 \ T
\sum_{n= - \infty}^{+ \infty} \left[ \sum_{m=0}^{\infty} \left( -{3 \over 4} \right)^m {\alpha_s^m \over m+2}
\sum_{k=0}^{\infty}c_k(m+2) \alpha_s^k \ln^k z_n \right],
\end{equation}
which shows explicitly that it is an $\alpha_s^2$-order term.

It is also convenient to present the integral (4.27) in the same way, namely

\begin{equation}
P'_{PT}(\Delta^2;T) = -{ 9 \over 2 \pi^2} \alpha^2_s \int_{\Lambda_{YM}}^{\infty} \dd\omega \ \omega^2 \ T
\sum_{n= - \infty}^{+ \infty} \left[ \sum_{m=0}^{\infty} \left( -{3 \over 4} \right)^m {\alpha_s^m \over m+2} P^{(n)}_{m+2}(\Delta^2) \sum_{k=0}^{\infty}c_k(m+2) \alpha_s^k \ln^k z_n \right]
\end{equation}
with

\begin{eqnarray}
P^{(n)}_{m+2}(\Delta^2) &=&
\sum_{p=0}^{m+1} {m+2 \choose p} M^{p-m-2}(\bar \omega^2, \omega^2_n) \left( - {3 \over 4} \Delta^2 \right)^{m+2-p}
\nonumber\\
&=&  \left( {3 \over 4} \Delta^2 \right)^2  M^{-2}(\bar \omega^2, \omega^2_n)  \sum_{p=0}^{m+1} {m+2 \choose p} M^{p-m}(\bar \omega^2, \omega^2_n) \left( - {3 \over 4} \Delta^2 \right)^{m-p} \nonumber\\
&=& \left( {3 \over 4} \Delta^2 \right)^2 P'^{(n)}_{m+2}(\Delta^2). \nonumber\\
\end{eqnarray}
Then the previous integral (4.30) becomes

\begin{eqnarray}
P'_{PT}(\Delta^2;T) = - \left( {9 \alpha_s \Delta^2 \over \sqrt2 4 \pi } \right)^2   \int_{\Lambda_{YM}}^{\infty} \dd\omega \ \omega^2 \ T
\sum_{n= - \infty}^{+ \infty} \left[ \sum_{m=0}^{\infty} {(-3/4)^m \alpha_s^m \over m+2} P'^{(n)}_{m+2}(\Delta^2) \sum_{k=0}^{\infty}c_k(m+2) \alpha_s^k \ln^k z_n \right],
\nonumber\\
\end{eqnarray}
and in the formal PT $\Delta^2=0$ limit it is zero as the whole expansion (4.13).
The integral (4.32) is really of the order of $\alpha_s^2$. However, this order numerically is much smaller than the corresponding order term in the expansion (4.13). This will be true for any corresponding orders in the expansions (4.13) and (4.32) because of the initial condition $x(\omega^2, \omega^2_n) \ll 1$ in Eq.~(4.10). Obviously, the structures of all expansions (4.13), (4.29) and (4.32) differ from each other.

\section{Convergence of the perturbative series }

The convergence of the power series (4.10) over $m$ is obvious, since it comes from the expansion of the logarithm (4.6). In turn, this means that the series in integer powers of $\alpha_s$ (4.29) and (4.32), due to the relation (4.25), are also convergent. At the same time, the convergence of the series (4.13) over $k$ is not so obvious and requires some additional investigation. The series (4.13) can be formally considered as a power series over small $\alpha_s$ with the coefficients $P_k(\Delta^2; T)$. The QCD fine-structure coupling constant correctly calculated at any scale is indeed always small. Let us remind that in this paper we use its value calculated at the $Z$-boson mass, namely $\alpha_s = \alpha_s(m_Z) = 0.1184$ \cite{24}. For the convergence of the power series (4.13) it is
necessary to show that its radius of convergence $r$ is bigger than any possible value of $\alpha_s$. The radius of convergence of the power series can be calculated in accordance with the Cauchy-Hadamard (CH) theorem as follows:

\begin{equation}
r^{-1} = \lim_{k \rightarrow \infty} \left|{ P_{k+1}(\Delta^2; T) \over P_k(\Delta^2; T)} \right|
\end{equation}
if this limit exists. Substituting the corresponding expressions from Eq.~(4.14), one obtains

\begin{equation}
r^{-1} = \lim_{k \rightarrow \infty} \left| { \int_{\Lambda_{YM}}^{\infty} \dd\omega \ \omega^2 \ \sum_{n= - \infty}^{+ \infty} \left[ { 1 \over M(\bar \omega^2, \omega^2_n)} \ln^k z_n \right] \over  \int_{\Lambda_{YM}}^{\infty} \dd\omega \ \omega^2 \ \sum_{n= - \infty}^{+ \infty} \left[ { 1 \over M(\bar \omega^2, \omega^2_n)} \ln^{k-1} z_n \right] } \right| \longrightarrow 1.
\end{equation}
Thus the series (4.13) converges absolutely for $\alpha_s < r=1$ and converges uniformly on every compact
subset of $\{ \alpha_s: \alpha_s < r \}$. Roughly speaking this means that any further calculated term
in integer powers of $\alpha_s$ will be smaller than the previous one (at least by one order of magnitude). Let us
also note that now there is no need in the restriction $|\alpha_s \ln z_n | < 1$, only $\alpha_s < 1$ is required, which always holds.

Summing up all the PT contributions, the PT pressure (2.5) finally becomes

\begin{equation}
P_{PT}(T) = P_{PT}(\Delta^2; T) + P'_{PT}( T) = [P_{PT}(\Delta^2; T) + P'_{PT}(\Delta^2; T)] + \tilde{P}_{PT}(T).
\end{equation}
Due to this decomposition a few things should be made perfectly clear. All three integrals (2.3), (2.4) and (2.5), contributing to the gluon pressure (2.2) are the corresponding parts of the initial skeleton loop integral derived
in \cite{15}. The first two integrals depend on the mass gap $\Delta^2$ only, so they are truly NP, indeed.
We call the integral (2.5) or, equivalently, (5.3) the PT one because it explicitly depends on $\alpha_s$, but this is only a convention.
Its two terms $P_{PT}(\Delta^2; T)$ (4.13) and $P'_{PT}(\Delta^2; T)$ (4.32) depend on the mass gap
$\Delta^2$ as well (they vanish in the PT $\Delta^2=0$ limit).
Its third term $\tilde{P}_{PT}(T)$
(4.29) begins with the $\alpha^2_s$-order correction, and does not depend on the the mass gap at all.
However, it is an infinite sum of pure PT contributions, i.e., in fact this is the corresponding part of the skeleton loop term (5.3).
The terms  $P_{PT}(\Delta^2; T)$ and $P'_{PT}(\Delta^2; T)$, being by themselves the two other corresponding parts
of the skeleton loop term (5.3), can be considered as the  $\alpha_s$- and $\alpha^2_s$-order corrections to the skeleton loop term $P_{YM}(T)$ (2.4), since they depend on the massive gluonic excitation $\bar \omega$. Let us emphasize that there cannot be any PT corrections to the bag constant $B_{YM}(T)$, since its skeleton part (2.3)
has been defined from the very beginning by the subtraction of all the types of the PT contributions \cite{14,15}.
All three expansions (4.13), (4.29) and (4.32) analytically depend on $\alpha_s$, and they can be calculated termwise in integer powers of $\alpha_s$.
The convergence of the series in integer powers of a small $\alpha_s$ derived here has been confirmed.
It guarantees that any further PT contribution in powers of $\alpha_s$ will be numerically smaller than the
previous one. It is worth emphasizing that in any case none of the PT contributions, and hence none of their sum, can be numerically bigger than the SB term, which describes the thermodynamic structure of the GM at high temperatures,
see Figs. 1 and 7 below.

Here a few general remarks are in order. In the initial thermal PT QCD the dependence on $\alpha_s$ is non-analytical, i.e., the expansion contains its fractional powers, $\alpha^3_s \ln \alpha_s$, etc. (see, for example \cite{5,12,17,26,27} and references therein). This leads to the divergent series in the thermal PT QCD, though each term up to the $\alpha^3_s \ln \alpha_s$-order has been calculated correctly. In \cite{15} it has briefly been explained why this effect occurs there. This problem does not occur in our formalism, which is the NP from the very beginning as underlined above.
Apparently, the reason is that it allows to calculate the $\alpha_s$-dependent corrections, which by themselves are
the corresponding skeleton parts of the skeleton loop integral (2.5), to
the NP term (2.4) which is the skeleton part of the initial skeleton loop integral (2.2).
In other words, the re-summation of an infinite number of the corresponding contributions has been already done for all of them. So our approach allows to develop the PT series in integer powers of $\alpha_s$ for
the NP quantities, i.e., the coefficients of these infinite series are the NP quantities.
Let us emphasize that these conventionally called PT series are, in fact, the so-called cluster expansions \cite{18}, which are convergent series and depend on the coupling constant not in elementary way, like all our expansions discussed here. This is completely different from the thermal PT QCD, which in the best case is of the asymptotic-type expansion. It is dealing with the pure PT contributions from the very beginning, and the necessary re-summation of the hard thermal loops (HTL) \cite{12} leads finally to the non-analytical dependence on $\alpha_s$. At the same time,
the formalism developed here makes it possible to calculate the $\alpha_s$-dependent contributions to the gluon pressure in terms of the convergent series in integer powers of a small $\alpha_s$. This seems to resolve the above-mentioned long-term problem.

However, instead it may produce another problem, namely the question of double-counting in
$\alpha_s$. In the present investigation there is no double-counting in $\alpha_s$, since all the convergent PT series (4.13), (4.29) and (4.32) are of different structure, as emphasized above. The series (4.29) and (4.32) begin with the
$\alpha_s^2$-orders, and therefore their numerical contributions are very small, since the series are convergent.
The convergent series (4.13) begins with the $\alpha_s$-order, which is the only one to be numerically calculated below. The double-counting problem will indeed arise when we will include the SB term into the full EoS, since it should be approached in the AF way, which involves all powers of $\alpha_s$. So the question how far do the pure PT corrections overlap with the NP ones cannot be answered here. But the problem is fixed, and we will address it and
clarify the situation in the forthcoming paper.

\section{The gluon pressure $P_g(T)$}

Summing up all contributions, the gluon pressure (2.2) thus finally becomes

\begin{equation}
P_g(T) = P_{NP}(T) + P_{PT}(T) = P_{NP}(T) + [P_{PT}(\Delta^2; T) + P'_{PT}(\Delta^2; T)] + \tilde{P}_{PT}(T).
\end{equation}
In general, both expansions $P_{PT}(\Delta^2; T)$ (4.13) and $P'_{PT}(\Delta^2; T)$ (4.32) are to be considered as producing the corresponding PT corrections to the leading NP part $P_{NP}(T)$ (3.1) of the GM EoS (6.1). At the same time, the pure PT term $\tilde{P}_{PT}(T)$ (4.29) is to be considered as producing the PT corrections to the leading PT contribution which is nothing but the above-mentioned SB term. However, due to the normalization condition of the free PT vacuum to zero (i.e., normalization of $P_g(T)$ to zero when the interaction is switched off), it is not
{\it explicitly} present in Eq.~(6.1) (see discussion below).
{\bf So none of pure PT corrections has to be calculated unless the leading SB term is restored to Eq.~(6.1) in a self-consistent way}. In this connection, let us note that the SB term or, equivalently, the pressure of a gas of massless free gluons (ideal gas) can be considered as the $\alpha_s^0=1$-order pure PT contribution to the full pressure.
That is why we start the numerical calculation of the PT contributions to the gluon pressure (6.1) from its first non-trivial order, namely the $\alpha_s$-order in the expansion (4.13). As emphasized above, it is the
$\alpha_s$-order contribution to $P_{YM}(T)$ in the NP term (3.1), which is already present in the GM EoS (6.1).
Then it looks like

\begin{equation}
P_g(T) = P_{NP}(T) + P^s_{PT}(T) + O(\alpha^2_s),
\end{equation}
where $P^s_{PT}(T) = \alpha_s P_1(\Delta^2; T)$ and for $P_1(\Delta^2; T)$ see Eq.~(4.16).
Omitting the terms of the $O(\alpha^2_s)$-order, for convenience, it is instructive to explicitly gather all our
results from the relations (3.1)-(3.3) and (4.15) for the gluon pressure (6.2) once more as follows:

\begin{equation}
P_g(T) = {6 \over \pi^2} \Delta^2 P_1 (T) + {16 \over \pi^2} T [P_2(T) + P_3(T) - P_4(T)] + P^s_{PT}(T),
\end{equation}
where the integral $P^s_{PT}(T)$ is

\begin{equation}
P^s_{PT}(T) =  \alpha_s \times {9 \over 2 \pi^2} \Delta^2
\int_{\Lambda_{YM}}^{\infty} \dd\omega \ \omega^2 \
{ 1 \over \bar \omega} { 1 \over e^{\beta \bar \omega}  - 1},
\end{equation}
while all other integrals $P_n(T), \ n=1,2,3,4$ are given in Eqs.~(3.2) and (3.3). This form is convenient for the numerical calculations. Let us note that when the interaction is formally switched off, i.e., letting $\alpha_s= \Delta^2=0$, the composition $[P_2(T)+P_3(T) -P_4(T)]$ becomes identical zero (see Eqs.~(3.3)) and thus
$P_g(T)$ itself. The gluon pressure (6.3) and its first PT contribution of the $\alpha_s$-order $P^s_{PT}(T)$ (6.4)
are also shown in Fig. 1. However, it is worth emphasizing once more that, in fact, the term $P^s_{PT}(T)$ is NP, depending on the mass gap $\Delta^2$, which is only suppressed by the $\alpha_s$-order.

\begin{figure}
\begin{center}
\includegraphics[width=10cm]{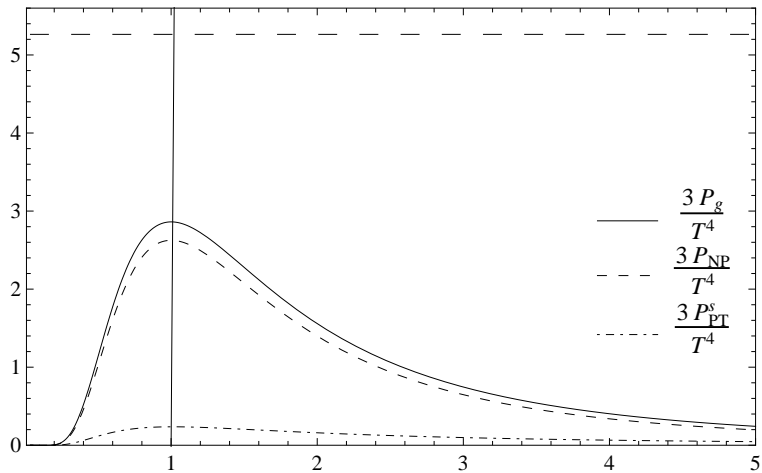}
\caption{The gluon pressure (6.3), the NP pressure (3.1) and the $\alpha_s$-dependent PT pressure (6.4), all
scaled (i.e., divided) by $T^4 / 3$, are shown as functions of $T/T_c$. Effectively, all curves have maxima at
$T_c = 266.5 \ \MeV$ (vertical solid line). The horizontal dashed line is the SB constant $3P_{SB}(T)/ T^4 =(24/45) \pi^2 \approx 5.26$.}
\label{fig:1}
\end{center}
\end{figure}

\begin{figure}
\begin{center}
\includegraphics[width=10cm]{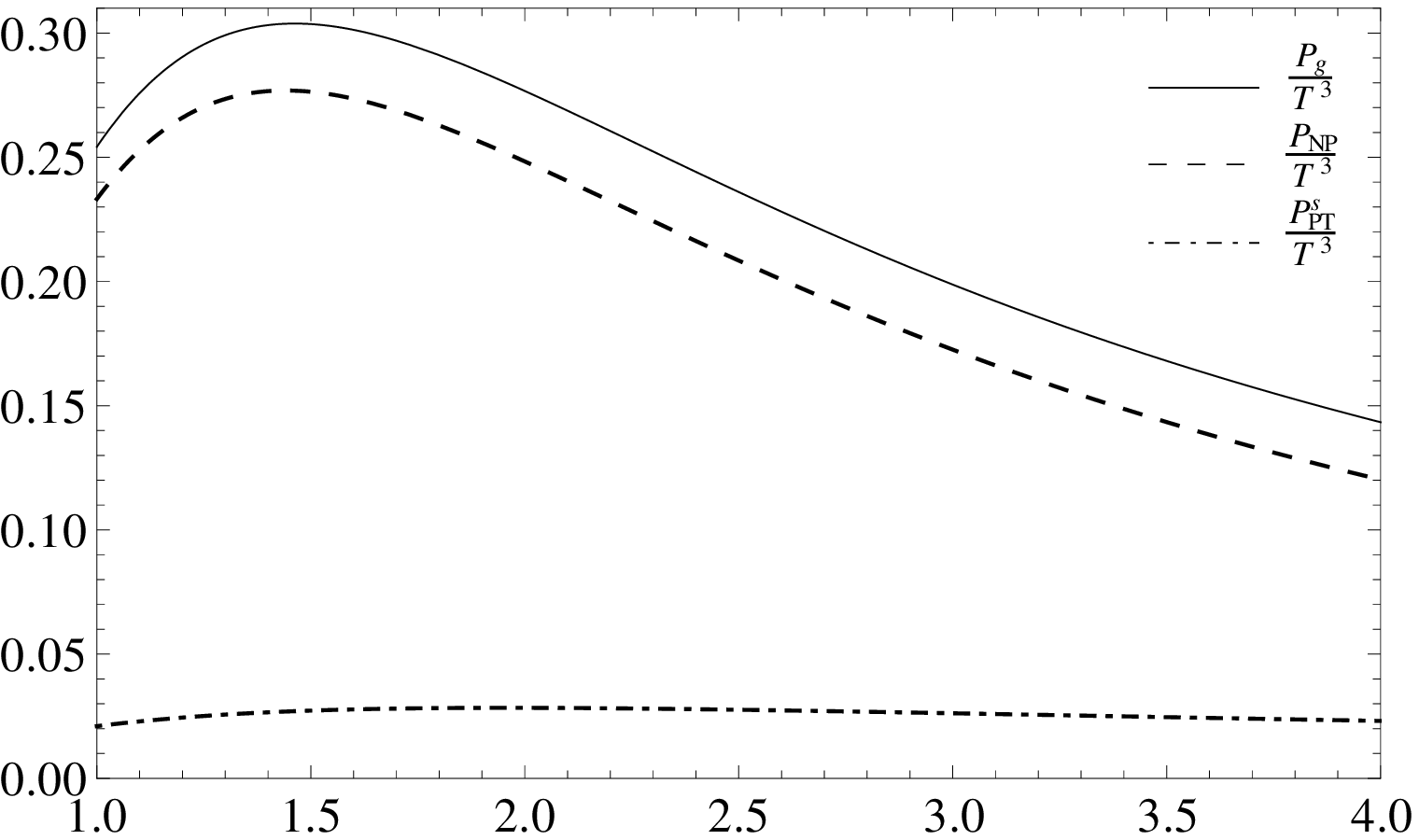}
\caption{The gluon pressure (6.3), the NP pressure (3.1)
and the $\alpha_s$-dependent PT pressure (6.4), all properly scaled in $\GeV$ units, are shown as functions
of $T/T_c$.}
\label{fig:2}
\end{center}
\end{figure}

\begin{figure}
\begin{center}
\includegraphics[width=10cm]{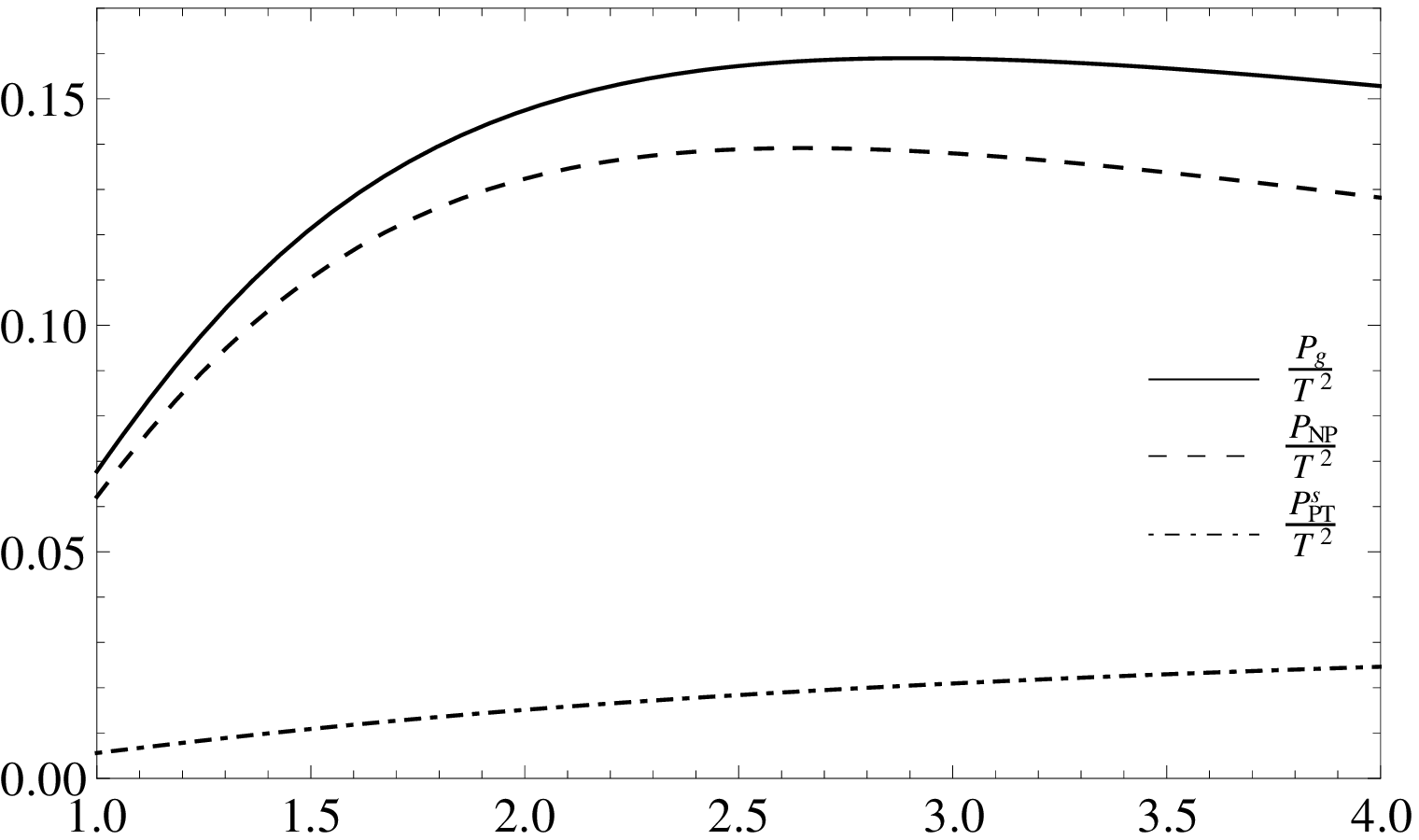}
\caption{The gluon pressure (6.3), the NP pressure (3.1)
and the $\alpha_s$-dependent PT pressure (6.4), all properly scaled in $\GeV^2$ units, are shown as functions of $T/T_c$.}
\label{fig:2}
\end{center}
\end{figure}

\begin{figure}
\begin{center}
\includegraphics[width=10cm]{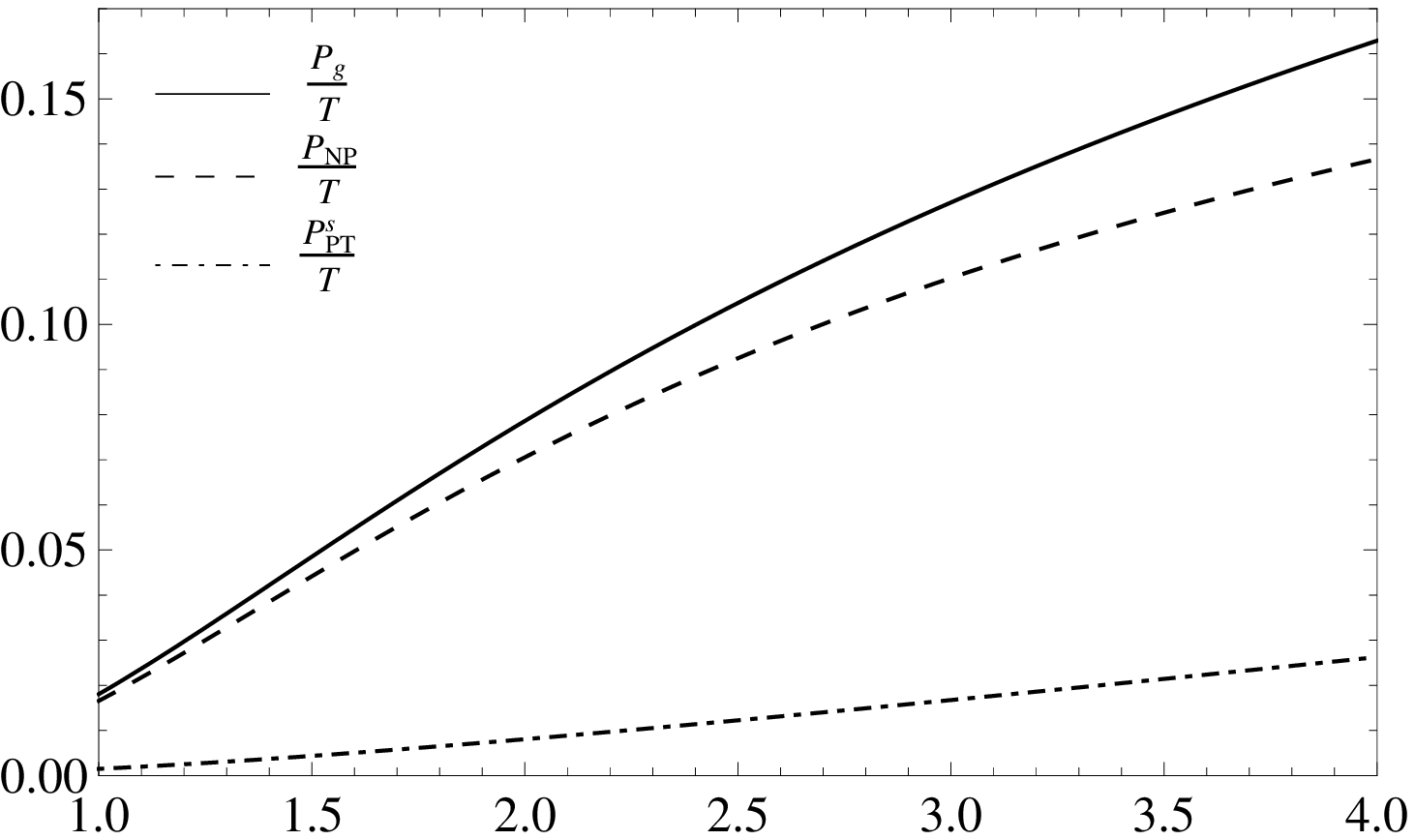}
\caption{The gluon pressure (6.3), the NP pressure (3.1)
and the $\alpha_s$-dependent PT pressure (6.4), all properly scaled in $\GeV^3$ units, are shown as functions of $T/T_c$. }
\label{fig:2}
\end{center}
\end{figure}

\begin{figure}
\begin{center}
\includegraphics[width=10cm]{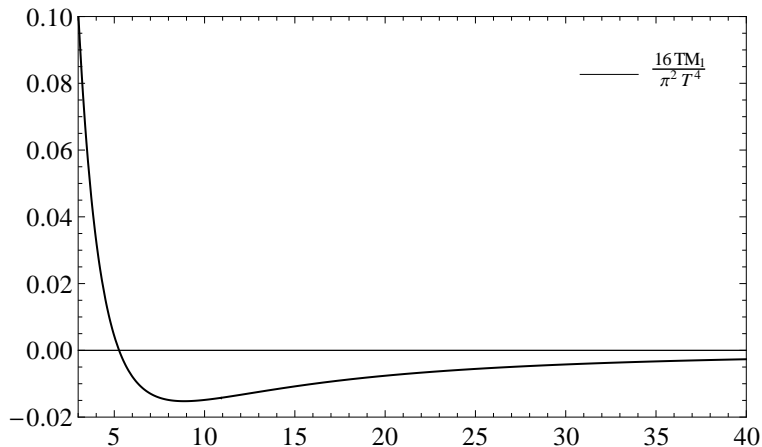}
\caption{The composition (B25) scaled by $T^4$ is shown as a function of $T/T_c$. It approaches
zero from below. This means that it does not contain the constant SB term, i.e., the SB terms have already been  exactly canceled far below $5T_c$. }
\label{fig:2}
\end{center}
\end{figure}

\begin{figure}
\begin{center}
\includegraphics[width=10cm]{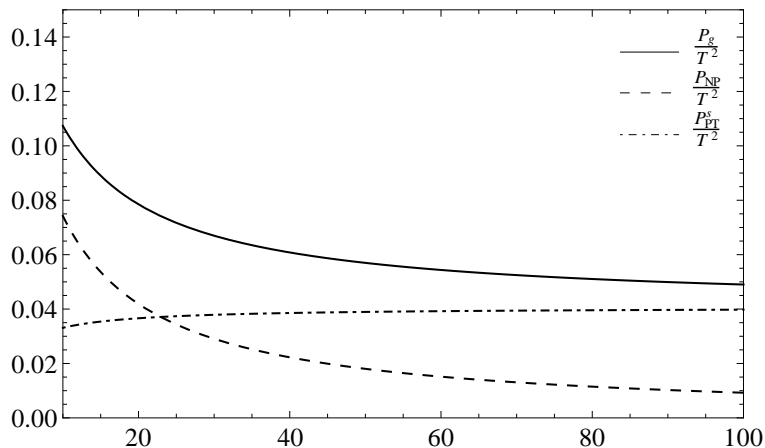}
\caption{The high temperature asymptotics of the gluon pressure (6.3), the NP pressure (3.1)
and the $\alpha_s$-dependent PT pressure (6.4) in $\GeV^2$ units are shown as functions of $T/T_c$.
At $T=23T_c$ the NP pressure $P_{NP}(T)$ goes below the PT pressure $P^s_{PT}(T)$.}
\label{fig:2}
\end{center}
\end{figure}

\begin{figure}
\begin{center}
\includegraphics[width=10cm]{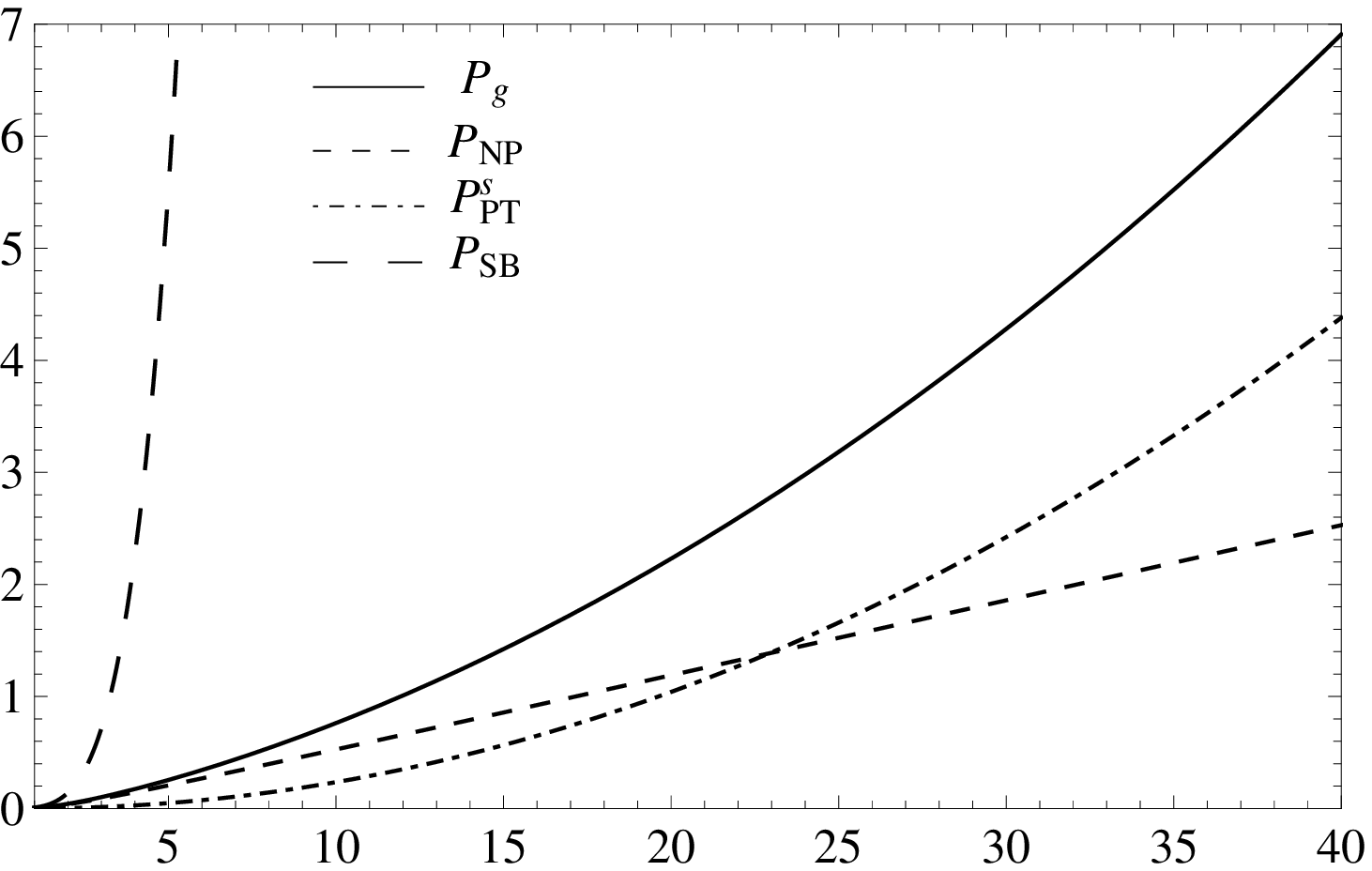}
\caption{The high temperature asymptotics of the gluon pressure (6.3), the NP pressure (3.1) and the $\alpha_s$-dependent PT pressure (6.4) in $\GeV^4$ are shown as functions of $T/T_c$. At $T=23T_c$ the NP pressure $P_{NP}$ goes below $P^s_{PT}(T)$. The SB pressure $P_{SB}(T) = (8/45) \pi^2 T^4$, formally extended up to zero temperature, is also shown.}
\label{fig:2}
\end{center}
\end{figure}

\section{Results, discussion and conclusions}

\subsection{Results}

From our numerical results it follows that all three pressures $P_g(T), \ P_{NP}(T)$ and $P^s_{PT}(T)$ as
functions of $T/T_c$ effectively have maxima at
the same "characteristic" temperature $T=T_c = 266.5 \ \MeV$, when they are scaled (i.e., divided) by $T^4/3$, see Fig. 1. From this figure it clearly follows that the gluon pressure (6.3) will never reach the SB constant $(24/45) \pi^2 = 3 P_{SB}(T)/T^4$ limit at high temperatures. That is not a surprise, since the SB term has been canceled in the gluon pressure from the very beginning due to the normalization condition of the free PT vacuum to zero \cite{13,14,15}
(see discussion below). On the other hand, from Fig. 1 it follows that all excitations of the NP dynamical origin, described by $P_{NP}(T)$ and $P^s_{PT}(T)$, and hence by $P_g(T)$ itself, do not survive in the very high temperature
$T \rightarrow \infty$ limit.
From Fig. 1 it is also clear that the first PT correction (6.4) is indeed smaller than the NP term (3.1) in the
moderately high temperature range up to approximately $(3-4)T_c$ \cite{15}.
At the maximum the PT term numerically is $0.236$, while the NP term is $2.63$, i.e., the former is by one order of magnitude smaller than the latter one. Due to the chosen scaling both pressures rather rapidly approach zero in the limit of very high temperatures.

It is worth noting here that the PT term (6.4) of the gluon pressure (6.3) describes the same massive gluonic excitations $\bar \omega$ (3.5) which have already been created by the NP dynamics in the GM.
However, they are suppressed by the $\alpha_s$-order, and therefore can be considered as new massive gluonic excitations in the GM. We will denote it as $\alpha_s \cdot \bar \omega$. So in total the gluon pressure (6.3) describes the five different massless and massive gluonic excitations within our approach.
In any case, the different types of the massive and massless gluonic excitations of the dynamical origin will necessary appear at non-zero temperature \cite{5,15,28,29,30} (and references therein).

Let us briefly discuss the asymptotic properties of all three pressures in more detail.
Below $T_c$ all pressures are exponentially suppressed in the $T \rightarrow 0$ limit, see Fig. 1. This is explicitly shown analytically in appendix A by considering the integrals (3.2)-(3.3) and (6.4) in the above-mentioned limit. Of course, this suppression is related to the low-temperature asymptotics of the gluon mean number (3.6), as mentioned above. The high-temperature ($T \rightarrow \infty$) expansion is explicitly evaluated in appendix B. At moderately high temperatures up to approximately $(3-4)T_c$ the exact functional dependence on $T$ remains rather complicated. It cannot be determined by the analytical evaluation of the above-mentioned integrals -- only numerically as shown in Fig. 1. This non-trivial $T$-dependence can also be seen in Figs. 2, 3 and 4.
In each of these figures all three pressures are scaled in the same way. Fig. 5 can be interpreted as clear
diagrammatic evidence of the exact cancelation of the SB terms close to $T_c$. It is analytically shown in Eq.~(B25), i.e., it occurs within the composition $16 T M_1(T)/ \pi^2T^4$. In the NP pressure (3.1) the exact cancelation of the mass gap terms $\Delta^2 T^2$ also occurs somewhere rather close to $T_c$, see Fig. 3. For the analytical evaluation of this phenomenon see appendix B, in general, and the high-temperature expansion (B24), in particular. As a result, the NP pressure (3.1) will scale as $T$, while the PT pressure (6.4), and hence the gluon pressure (6.3), will continue to scale up to the leading order as $\sim \alpha_s \Delta^2T^2$ in Eqs.~(B33) and (B34), respectively. Thus both pressures will approach the same constant in the limit of high temperature in Fig. 6, but very slowly since the term $\sim T^2$ is suppressed by the $\alpha_s$ order. At $T=23T_c$ the NP pressure goes below
the PT one, see Figs. 6 and 7. In general, all pressures are polynomials in integer powers of $T$ up to $T^2$ at very high temperatures. The term $\sim T^2$ has been first introduced in the phenomenological EoS \cite{31} (see also \cite{32,33,34,35,36} and references therein). On the contrary, in our approach both terms $\sim T^2$ and $\sim T$ have not been introduced by hand. They naturally appear as a result of the explicit presence of the mass gap from the very beginning in the NP analytical EoS \cite{15}.

\subsection{Conclusions}

Our final conclusions are as follows:

{\bf (i)}. The effective potential for composite operators \cite{13} may provide a new general analytic
approach to QCD at non-zero temperature and density \cite{14,15}.

{\bf (ii)}. It is essentially NP by origin, but may incorporate the thermal PT expansion.

{\bf (iii)}. We have developed the analytic thermal PT in the form of the convergent series, which made it possible to calculate the PT part of the gluon pressure termwise in integer powers of a small $\alpha_s$.

{\bf (iv)}. We have shown that the PT contribution of the $\alpha_s$-order is numerically much smaller than the NP term in the range up to $23T_c$, see Figs. 1-4 and 6-7.

{\bf (v)}. In the gluon pressure (6.1) the higher order terms in integer powers of a small $\alpha_s$, which are determined by the convergent expansions (4.29) and (4.32), can be neglected.

{\bf (vi)}. All three pressures scaled (i.e., divided) by $T^4/3$ have maxima at the temperature $T=T_c= 266.5 \ \MeV$. Their low- (below $T_c$) and high-temperature (above $T_c$) expansions have been evaluated in appendices A and B, respectively.

{\bf (vii)}. In the low-temperature $T \rightarrow 0$ limit all three pressures are exponentially
suppressed (Fig. 1)
due to the corresponding asymptotic of the gluon mean number. In the low-temperature expansion of the NP part (see Eq.~{A28)) a non-analytical dependence on the mass gap appears in terms
$\sim (\Delta^2)^{1/2} T^3 \sim \Delta T^3$.

{\bf (viii)}. The complicated mass gap- and $T$-dependence of all three pressures near $T_c$ and up to
approximately $(3-4)T_c$ is seen in Figs. 1, 2, 3, 4. In particular, the behavior of each pressure is not exactly
power-type. This especially clearly follows from Figs. 2, 3 and 4.

{\bf (ix)}. The polynomial character of the high-temperature expansions for all three pressures is confirmed due to
the corresponding asymptotic of the gluon mean number.

\hspace{10mm} {\bf (a)}. For the NP pressure (3.1) it contains only terms $\sim T$, and some of them

\hspace{10mm}  depend non-analytically on the mass gap, namely $\sim (\Delta^2)^{3/2} T \sim \Delta^3 T$.

\hspace{10mm} {\bf (b)}. For the $\alpha_s$-dependent PT contribution (6.4) it contains the terms $\sim T^2$

\hspace{10mm}  and $\sim T$ with a non-analytical dependence on the mass gap as above.

\hspace{10mm} {\bf (c)}. For the gluon pressure (6.3) it contains both types of terms.

{\bf (x)}. In low- and high-temperature expansions, a non-analytical dependence on the mass gap occurs, as described above, but it is not an expansion parameter like is $\alpha_s$.

{\bf (xi)}. In the gluon pressure (B34) the mass gap term $\sim \Delta^2 T^2$ is explicitly present in the whole temperature range, though it is suppressed by the $\alpha_s$-order.

{\bf (xii)}. The PT part (B26) dominates over its NP counterpart (B1) in the limit of very high temperature
starting from $T=23T_c$, see Figs. 6 and 7. It is expected from the general point of view. This underlines
once more the importance of the calculated here the $\alpha_s$-dependent PT pressure.

{\bf (xiii)}. Our analytical derivations in appendices A and B are in complete agreement with our numerical results
shown in Figs. 1-7 and vice-versa.

{\bf (xiv)}. In addition to the two massless $\omega_1, \ \omega_2$ and the two massive $\omega', \ \bar \omega$ gluonic excitations described by $P_{NP}(T)$ (3.1), we have also one new massive excitation $\alpha_s \cdot \bar \omega$ described by $P_{PT}^s (T)$ (6.4). All these excitations are of the NP, dynamical origin,
since in the PT $\Delta^2=0$ limit they vanish from the spectrum.

\vspace{3mm}

The NP part (3.1) determines the thermodynamic structure of the GM at low temperatures within our approach. It is uniquely fixed and numerically it is half of the SB value at $T_c$ (Fig. 1 and \cite{15}).
The main problem which remains to solve is how to include the massless free gluons contribution to the gluon pressure (6.3), in order to reach the SB limit at high temperatures in the GM. Above $T_c$ it will be always much bigger than any other correctly calculated PT contributions to the gluon pressure, see Figs. 1 and 7.
The above-mentioned problem is not an easy task due to the normalization condition of the free PT vacuum to zero. The SB term cannot simply be added to the gluon pressure even multiplied by the corresponding
$\Theta ( (T / T_c) - 1)$-function. This will lead to the finite jump at $T=T_c$ in the full pressure which is not acceptable. The full pressure is always a continuous growing function of temperature at any point of its domain. So its inclusion into the YM EoS (which describes the thermodynamic structure of the GM in the whole temperature range) should be done in a more sophisticated way (work is in progress). Only after completing this program we will be able to analyze and compare our results for $SU(3)$ GM EoS with the lattice results in \cite{32,33,34,37,38,39} and analytical ones in \cite{5,27,29,30,31,35,36} (and references therein). The comparison and agreement with the thermal lattice calculations is especially important, since it will make it possible to understand what is the physics behind them.
Our ultimate goal is the description of the dynamical structure of the QGP within NP analytical approach to QCD
at non-zero temperature and density developed and advocated here and in \cite{14,15}.

However, let us underline in advance that the gluon pressure $P_g(T)$ analytically investigated and numerically calculated in the present investigation will be the most important part of the full pressure. Just it will be mainly responsible for its NP physics and will provide the agreement with lattice data in the whole temperature range and especially near $T_c$.
The above-mentioned addition of the SB term will only ensure the correct high temperature limit of the full pressure.
And finally, we do not expect that the inclusion of quark degrees of freedom will somehow affect the convergence of the corresponding PT series. But this has to be investigated and checked after $SU(3)$ YM EoS is fixed, i.e., when the above-mentioned program is completed.

\begin{acknowledgments}

This paper is dedicated to the memory of Prof. J. Zim\'anyi, who has initiated this investigation.
Support by the Hungarian National Fund OTKA - 77816 (P. L\'evai) is to be acknowledged.
We thank R. Pisarski for bringing our attention to the reference \cite{31}.
We would like also to thank T. Bir\'{o}, T. Csorg\"{o}, P. V\'an, G. Barnaf\"{o}ldi,
A. Luk\'acs, J. R\'evai, A. Shurgaia, J. Nyiri and S. Pochybova for useful discussions, remarks and help.
One of the authors (V.G.) is grateful to V.K. and A.V. Kouzushins, F. Todua, R. Shakarishvili, I. Kiguradze, N. Partsvania, M. Eliashvili for constant support, help and interest.

\end{acknowledgments}

\appendix

\section{Low-temperature expansion}

Let us begin with noting in advance that all exactly calculated integrals, discussed in appendices A and B,
can be found in \cite{25,40}. This is also true for the asymptotics in the low- and high-temperature limits of those integrals which cannot be analytically derived. In order to evaluate a low-temperature expansion for the gluon pressure

\begin{equation}
P_g(T) = P_{NP}(T) + P^s_{PT}(T),
\end{equation}
it is convenient to present the NP pressure as in Eq.~(3.1), namely

\begin{equation}
P_{NP}(T) = {6 \over \pi^2} \Delta^2 P_1 (T) + {16 \over \pi^2} T M(T),
\end{equation}
where the integral $P_1(T)$ (3.2), on account of Eq.~(3.6), is

\begin{equation}
P_1(T) = \int_{\omega_{eff}}^{\infty} \dd \omega \ \omega \ N_g(\beta, \omega) =
\int_{\omega_{eff}}^{\infty} \dd \omega {\omega \over e^{\beta\omega} -1},
\end{equation}
and

\begin{equation}
M(T) = P_2(T) + P_3(T) - P_4(T),
\end{equation}
with

\begin{eqnarray}
P_2(T) &=& \int_{\omega_{eff}}^{\infty} \dd \omega \ \omega^2
\ln \left( 1- e^{-\beta\omega} \right), \nonumber\\
P_3(T)&=& \int_0^{\omega_{eff}} \dd \omega \ \omega^2 \ln
\left( 1 - e^{- \beta\omega'} \right), \nonumber\\
P_4(T) &=& \int_0^{\infty} \dd \omega \ \omega^2 \ln \left( 1 -
e^{- \beta \bar \omega} \right).
\end{eqnarray}

In all the above-displayed integrals the variable $y = e^{-\beta\omega}$ is always small, and hence
$y^{-1} = e^{\beta\omega}$ is always big,
in the low-temperature limit $T \rightarrow 0 \ (\beta = T^{-1} \rightarrow \infty)$. This is true for the
exponents $e^{-\beta\omega'}$ and $e^{-\beta \bar \omega}$ as well.
Then the gluon mean number $N_g(\beta, \omega)$ in the integral (A3) can be approximated as
$ N_g(\beta, \omega) \sim \exp(-\beta\omega)$ up to the leading order. So $P_1(T)$ becomes

\begin{equation}
P_1(T) = \int_{\omega_{eff}}^{\infty} \dd \omega \ \omega \ N_g(\beta, \omega) \sim
\int_{\omega_{eff}}^{\infty} \dd \omega \ \omega e^{- \beta\omega}.
\end{equation}
The almost trivial integration yields

\begin{equation}
P_1(T) \sim ( T^2 + \omega_{eff} T) e^{ - {\omega_{eff} \over T }}, \quad T \rightarrow 0.
\end{equation}

The integral $P_2(T)$ can be considered in the same way. Up to the leading order it becomes

\begin{equation}
P_2(T) = \int_{\omega_{eff}}^{\infty} \dd \omega \ \omega^2
\ln \left( 1- e^{-\beta\omega} \right) \sim - \int_{\omega_{eff}}^{\infty} \dd \omega \ \omega^2
\ e^{-\beta\omega}, \quad \beta \rightarrow \infty,
\end{equation}
and integrating it, one obtains

\begin{equation}
P_2(T) \sim - ( 2T^3 + 2 \omega_{eff} T^2 + \omega^2_{eff} T) e^{ - {\omega_{eff} \over T }}, \quad T \rightarrow 0.
\end{equation}

The integral $P_3(T)$ up to the leading order looks like

\begin{equation}
P_3(T)= \int_0^{\omega_{eff}} \dd \omega \ \omega^2 \ln
\left( 1 - e^{- \beta\omega'} \right) \sim - \int_0^{\omega_{eff}} \dd \omega \ \omega^2
e^{- \beta\omega'}, \quad \beta \rightarrow \infty,
\end{equation}
and replacing the variable $\omega$ by the variable $\omega'$ in accordance with the relation (3.4), this integral becomes

\begin{equation}
P_3(T) \sim - \int_a^{\omega'_{eff}} \dd \omega' \ \omega'
\sqrt{(\omega'^2 - a^2)} \ e^{- \beta\omega'}, \quad \beta \rightarrow \infty,
\end{equation}
where

\begin{equation}
\omega'_{eff} = \sqrt{(\omega_{eff}^2 + a^2)}, \ \quad a = \sqrt{3} \Delta.
\end{equation}
Unfortunately, this asymptotical expression (A11) cannot be directly evaluated, like it has
been done for the integrals (A6) and (A8). However, noting that the variable
$x=a^2 / \omega'^2 \leq 1$, we can formally expand

\begin{equation}
\sqrt{(\omega'^2 - a^2)}= \omega'( 1 - x)^{1/2} = \omega' \left[ 1 - {1 \over 2} {a^2 \over \omega'^2} + \sum_{k=2}^{\infty}{1/2  \choose k}(-x)^k \right].
\end{equation}
Then from the integral (A11) one obtains

\begin{equation}
P_3(T) \sim - \int_{\sqrt{3} \Delta }^{\omega'_{eff}} \dd \omega' \ \omega'^2 \ e^{- \beta\omega'}
+ {3 \over 2} \Delta^2 \int_{\sqrt{3} \Delta}^{\omega'_{eff}} \dd \omega' \ e^{- \beta\omega'} +
P^{(k)}_3(T), \quad \beta \rightarrow \infty,
\end{equation}
where

\begin{equation}
P^{(k)}_3(T) = - \int_{\sqrt{3} \Delta }^{\omega'_{eff}} \dd \omega' \ \omega'^2 \ e^{- \beta\omega'}
\sum_{k=2}^{\infty}{1/2  \choose k}(-x)^k.
\end{equation}
Let us consider the last integral (A15) in more detail. Since the series over $k$ are convergent in the interval
of integration and the functions depending on $k$ are integrable in this interval, these series may be integrated termwise \cite{25}, that is,

\begin{equation}
P^{(k)}_3(T) = - \sum_{k=2}^{\infty}{1/2  \choose k}(-a^2)^k \int_{\sqrt{3} \Delta }^{\omega'_{eff}} \dd \omega'
\ { e^{- \beta\omega'} \over (\omega')^{2k-2}}.
\end{equation}
Integrating it, one obtains

\begin{equation}
P^{(k)}_3(T) = - \sum_{k=2}^{\infty}{1/2  \choose k}(-a^2)^k \left[ N_3^{(k)}(T, \omega') \right]_{\sqrt{3} \Delta }^{\omega'_{eff}}
\end{equation}
and $\left[ N_3^{(k)}(T, \omega') \right]_{\sqrt{3} \Delta }^{\omega'_{eff}}$ denotes the result of the integration
over $\omega'$ in Eq.~(A16) on the interval $[\sqrt{3} \Delta, \omega'_{eff}]$, while the function $N_3^{(k)}(T, \omega')$ itself is

\begin{equation}
N^{(k)}_3(T, \omega') = - e^{- \beta\omega'} \sum_{m=1}^{2k-3}{ (- \beta)^{m-1} (\omega')^{m +2 - 2k}
\over (2k-3)(2k-4)...(2k-2-m)} + { (- \beta)^{2k-3} \over (2k-3)!} {\rm Ei}(- \beta\omega').
\end{equation}
The series for the exponential integral function ${\rm Ei}(- \beta\omega')$ is \cite{25}

\begin{equation}
{\rm Ei}(- \beta\omega') =  e^{- \beta\omega'} \sum_{l=1}^n (-1)^l { (l-1)! \over (\beta\omega')^l},
\quad n \geq 2k-3, \quad \beta \rightarrow \infty.
\end{equation}
If one chooses $n = 2k-3$ in the previous equation (i.e., neglecting the relative error of the approximation (A19)), it is easy to show that both terms in Eq.~(A18) for $N^{(k)}_3(T, \omega')$ cancel each other termwise for any $k \geq 2$, and thus

\begin{equation}
N^{(k)}_3(T, \omega') = 0, \quad k=2,3,4,... \
\end{equation}
or, equivalently,

\begin{equation}
P^{(k)}_3(T) = 0, \quad k=2,3,4,... \ .
\end{equation}

Going back to Eq.~(A14) and easily integrating the first two terms,
and taking into account the previous result, one comes to the following expansion

\begin{eqnarray}
P_3(T) &\sim& ( 2T^3 + 2 \omega'_{eff} T^2 + \omega'^2_{eff} T) e^{ - {\omega'_{eff} \over T }}
-( 2T^3 + 2 \sqrt{3} \Delta   T^2 + 3 \Delta^2 T) e^{ - { \sqrt{3} \Delta   \over T }} \nonumber\\
&-& { 3 \over 2} \Delta^2 T e^{ - {\omega'_{eff} \over T }}
+ { 3 \over 2} \Delta^2 T e^{ - { \sqrt{3} \Delta  \over T }}, \quad T \rightarrow 0.
\end{eqnarray}

The integral $P_4(T)$ up to the leading order looks like

\begin{equation}
P_4(T) = \int_0^{\infty} \dd \omega \ \omega^2 \ln \left( 1 -
e^{- \beta \bar \omega} \right) \sim - \int_0^{\infty} \dd \omega \ \omega^2
e^{- \beta \bar \omega},  \quad \beta \rightarrow \infty,
\end{equation}
and replacing the variable $\omega$ by the variable $\bar \omega$ in accordance with the relation (4.5),
this integral becomes

\begin{equation}
P_4(T) \sim - \int_{(a/2)}^{\infty} \dd \bar \omega \ \bar \omega
\sqrt{(\bar \omega^2 - (a/2)^2)} \ e^{- \beta \bar \omega}, \quad \beta \rightarrow \infty.
\end{equation}
Noting again that the variable $ z= a^2 / 4 \bar \omega^2 \leq 1$, we can formally expand

\begin{equation}
\sqrt{(\bar \omega^2 - (a/2)^2)}= \bar \omega \left[ 1 - (a^2 / 8 \bar \omega^2) + \sum_{k=2}^{\infty}{1/2  \choose k}(-z)^k \right].
\end{equation}
Then from the integral (A24) one obtains

\begin{equation}
P_4(T) \sim - \int_{\sqrt{3}\Delta/2 }^{\infty} \dd \bar \omega \ \bar \omega^2 \ e^{- \beta \bar \omega}
+ { 3 \over 8} \Delta^2 \int_{\sqrt{3}\Delta/2}^{\infty} \dd \bar \omega \ e^{- \beta \bar \omega} + P_4^{(k)}(T), \quad \beta \rightarrow \infty,
\end{equation}
where due to the same formalism which has been used previously in order to get the result (A21), one can conclude
that $P_4^{(k)}(T)=0$ as well. Easily integrating the first two terms, one comes to the following expansion

\begin{equation}
P_4(T) \sim - ( 2T^3 + \sqrt{3} \Delta T^2 + {3 \over 4} \Delta^2 T) e^{ - { \sqrt{3} \Delta \over 2 T }}
+ {3 \over 8} \Delta^2 T e^{- {\sqrt{3} \Delta \over 2T }}, \quad T \rightarrow 0.
\end{equation}

Substituting all these expansions into the Eq.~(A2), one obtains

\begin{eqnarray}
P_{NP}(T) &\sim& {6 \over \pi^2} \Delta^2 ( T^2 + \omega_{eff} T) e^{ - {\omega_{eff} \over T }}
- {16 \over \pi^2} T \left[ 2T^3 + 2 \omega_{eff} T^2 + \omega^2_{eff} T \right] e^{ - {\omega_{eff} \over T }}
\nonumber\\
&+& {16 \over \pi^2} T \left[ ( 2T^3 + 2 \omega'_{eff} T^2 + \omega'^2_{eff} T) e^{ - {\omega'_{eff} \over T }}
-( 2T^3 + 2 \sqrt{3} \Delta   T^2 + 3 \Delta^2 T) e^{ - { \sqrt{3} \Delta   \over T }}                                  \right] \nonumber\\
&-& {24 \over \pi^2} T^2 \Delta^2 \left[ e^{ - {\omega'_{eff} \over T }} - e^{ - { \sqrt{3} \Delta \over T }} \right]
+ {16 \over \pi^2} T \left[ 2T^3 + \sqrt{3} \Delta T^2 +
{3 \over 8} \Delta^2 T \right] e^{ - { \sqrt{3} \Delta \over 2 T }}, \ T \rightarrow 0. \nonumber\\
\end{eqnarray}
Evidently, this is nothing but a low-temperature expansion for the NP pressure $P_{NP}(T)$. Let us note that it contains a non-analytical dependence on the mass gap squared in terms $\sim (\Delta^2)^{1/2} T^3 \sim \Delta T^3$,
but the mass gap is not an expansion parameter like $\alpha_s$.

It is instructive to re-write this expansion as follows:

\begin{eqnarray}
P_{NP}(T) &\sim& F_{NP}(\Delta, T) + b_1 \left[- e^{ - {\omega_{eff} \over T }} + e^{ - {\omega'_{eff} \over T }}       - e^{ - { \sqrt{3} \Delta \over T }} + e^{ - { \sqrt{3} \Delta \over 2 T }} \right] P_{SB}(T) \nonumber\\
&-& {16 \over \pi^2} T^2 \left[ 2 \omega_{eff} T + \omega^2_{eff} \right] e^{ - {\omega_{eff} \over T }}
+ {16 \over \pi^2} T^2 \left[ 2 \omega'_{eff} T + \omega'^2_{eff} \right] e^{ - {\omega'_{eff} \over T }},
\ T \rightarrow 0, \nonumber\\
\end{eqnarray}
since for YM fields (see, for example \cite{11,12} and references therein)

\begin{equation}
P_{SB}(T) = {8 \over 45} \pi^2 T^4.
\end{equation}
$F_{NP}(\Delta, T)$ denotes the sum of all the terms which directly depend on the mass gap, so that $F_{NP}(T, \Delta =0)=0$ and $b_1=(180 / \pi^4)$. So the propagation of massless gluons below $T_c$ can be described by the
SB-type term. However, it is exponentially suppressed in the
$T \rightarrow 0$ limit, as it should be. At $T \sim T_c$ its contribution can be numerically comparable with other contributions in Eq.~(A28). That's no surprise that the massless gluons may be present in the GM at any temperature.
Moreover, let us note in advance that the propagation of the massless free gluons below $T_c$ (if any) cannot be described by the SB term itself. It should also be exponentially suppressed in the same way as it is shown in Eq.~(A29). But it has to survive in the PT $\Delta^2=0$ limit, while the contribution (A29) vanishes in this limit
($\omega'_{eff}= \omega_{eff}$
in this case, see Eq.~(A12)). The SB term can describe the propagation of the massless free gluons only in the high temperatures limit above $T_c$ (see appendix B below).

Let us now consider Eq.~(6.4), which in the $ T \rightarrow 0 \ (\beta = T^{-1} \rightarrow \infty)$ limit
up to the leading order becomes

\begin{equation}
P^s_{PT}(T) = {9 \alpha_s \over 2 \pi^2} \Delta^2
\int_{\Lambda_{YM}}^{\infty} \dd\omega \ \omega^2 \
{ 1 \over \bar \omega} { 1 \over e^{\beta \bar \omega}  - 1} \sim {9 \alpha_s \over 2 \pi^2} \Delta^2
\int_{\Lambda_{YM}}^{\infty} \dd\omega \ \omega^2 \
{ 1 \over \bar \omega}  e^{- \beta \bar \omega}, \ \beta \rightarrow \infty,
\end{equation}
and $\bar \omega$ is given by the relation (3.5). Replacing the variable $\omega$ by the variable
$\bar \omega$, as in Eq.~(A23), one obtains

\begin{equation}
P^s_{PT} (T) \sim {9 \alpha_s \over 2 \pi^2} \Delta^2 \int_{\tilde{\omega}_{eff}}^{\infty} \dd \bar \omega \
\sqrt{(\bar \omega^2 - (a/2)^2)} \ e^{- \beta \bar \omega}, \quad \beta \rightarrow \infty,
\end{equation}
where $\tilde{\omega}_{eff} = \sqrt{\Lambda^2_{YM} + (a/2)^2}$, and for $a$ see Eq.~(A12).
Noting that the variable $ z= a^2 / 4 \bar \omega^2 < 1$ in this case, we can use the expansion (A25) in order to obtain

\begin{equation}
P^s_{PT}(T) \sim  {9 \alpha_s \over 2 \pi^2} \Delta^2 \left[ \int_{\tilde{\omega}_{eff} }^{\infty} \dd \bar \omega \ \bar \omega \ e^{- \beta \bar \omega}
 -{ 3 \over 8} \Delta^2 \int_{\tilde{\omega}_{eff}}^{\infty} \dd \bar \omega \ {e^{- \beta \bar \omega} \over \bar \omega} + P_s^{(k)}(T) \right], \quad \beta \rightarrow \infty.
\end{equation}
Due to the same formalism which has been used previously in order to get the result (A21), one can conclude
that $P_s^{(k)}(T)=0$ as well. Easily integrating the first two terms, one comes to the following expansion

\begin{equation}
P^s_{PT}(T) \sim  {9 \alpha_s \over 2 \pi^2} \Delta^2 \left[ (T^2 + T \tilde{\omega}_{eff}) e^{- {\tilde{\omega}_{eff} \over T}} + {3 \over 8} \Delta^2 {\rm Ei}( - {\tilde{\omega}_{eff} \over T}) \right], \quad T \rightarrow 0.
\end{equation}
Here and below the corresponding exponential integral functions are defined by Eq.~(A19).
Summing up the expansions (A28) and (A34), one obtains a low-temperature expansion for the gluon pressure (A1)
as follows:

\begin{eqnarray}
\hspace{12mm} P_g(T) &\sim& {6 \over \pi^2} \Delta^2 ( T^2 + \omega_{eff} T) e^{ - {\omega_{eff} \over T }}
- {16 \over \pi^2} T \left[ 2T^3 + 2 \omega_{eff} T^2 + \omega^2_{eff} T \right] e^{ - {\omega_{eff} \over T }}
\nonumber\\
&+& {16 \over \pi^2} T \left[ ( 2T^3 + 2 \omega'_{eff} T^2 + \omega'^2_{eff} T) e^{ - {\omega'_{eff} \over T }}
-( 2T^3 + 2 \sqrt{3} \Delta   T^2 + 3 \Delta^2 T) e^{ - { \sqrt{3} \Delta   \over T }}                                  \right] \nonumber\\
&-& {24 \over \pi^2} T^2 \Delta^2 \left[ e^{ - {\omega'_{eff} \over T }} - e^{ - { \sqrt{3} \Delta \over T }} \right]
+ {16 \over \pi^2} T \left[ 2T^3 + \sqrt{3} \Delta T^2 +
{3 \over 8} \Delta^2 T \right] e^{ - { \sqrt{3} \Delta \over 2 T }} \nonumber\\
&+& {9 \alpha_s \over 2 \pi^2} \Delta^2 \left[ (T^2 + T \tilde{\omega}_{eff}) e^{- {\tilde{\omega}_{eff} \over T}} + {3 \over 8} \Delta^2 {\rm Ei}( - {\tilde{\omega}_{eff} \over T}) \right], \quad T \rightarrow 0.
\end{eqnarray}
Let us note that the low-temperature expansion (A35) depends mainly on the effective massive "excitation" $\omega'_{eff} = \sqrt{\omega^2_{eff} + a^2}$, and does not depend on the effective massive "excitation" $\bar \omega_{eff} = \sqrt{\omega^2_{eff} + (a/2)^2}$ at all, while the dependence on the effective massive "excitation" $\tilde{\omega}_{eff} = \sqrt{\Lambda^2_{YM} + (a/2)^2}$ is suppressed by $\alpha_s$.

It is instructive to use in the exponents of the previous expansion the following obvious relations: $\omega_{eff} = \nu_1 T_c, \ \omega'_{eff} = \nu_2 T_c, \ \sqrt{3} \Delta = \nu_3 T_c, \  \nu_4 = (1/2) \nu_3, \ \tilde{\omega}_{eff} = \nu_5 T_c$, since all numerical values of these parameters are known. Then the previous expansion looks like

\begin{eqnarray}
\hspace{12mm} P_g(T) &\sim& {6 \over \pi^2} \Delta^2 ( T^2 + \omega_{eff} T) e^{ -  \nu_1{T_c \over T }}
- {16 \over \pi^2} T \left[ 2T^3 + 2 \omega_{eff} T^2 + \omega^2_{eff} T \right] e^{ - \nu_1{T_c \over T }}
\nonumber\\
&+& {16 \over \pi^2} T \left[ ( 2T^3 + 2 \omega'_{eff} T^2 + \omega'^2_{eff} T) e^{ - \nu_2 {T_c \over T }}
-( 2T^3 + 2 \sqrt{3} \Delta   T^2 + 3 \Delta^2 T) e^{ - \nu_3 { T_c \over T }}                                  \right] \nonumber\\
&-& {24 \over \pi^2} T^2 \Delta^2 \left[ e^{ - \nu_2 {T_c \over T }} - e^{ - \nu_3{ T_c \over T }} \right]
+ {16 \over \pi^2} T \left[ 2T^3 + \sqrt{3} \Delta T^2 +
{3 \over 8} \Delta^2 T \right] e^{ - \nu_4 { T_c \over T }} \nonumber\\
&+& {9 \alpha_s \over 2 \pi^2} \Delta^2 \left[ (T^2 + T \tilde{\omega}_{eff}) e^{- \nu_5 {T_c \over T}} + {3 \over 8} \Delta^2 {\rm Ei}( - \nu_5 {T_c \over T}) \right], \quad T \rightarrow 0.
\end{eqnarray}
The expansion (A36) clearly shows that the exponential suppression of any pressure at low temperature
below $T_c$ is determined by the corresponding asymptotic of the gluon mean number (4.6), namely

\begin{equation}
N_g = {1 \over e^{\omega \over T} -1} \sim e^{- \nu { T_c \over T}}, \quad T_c > T \rightarrow 0,
\end{equation}
by replacing $\omega$ by $ \nu T_c$ in each different case, as it is seen in the previous low-temperature expansion
for the gluon pressure. For the scaled gluon pressure $3 P_g(T) / T^4$ the expansion (A36) is especially useful, since it depends on the dimensionless variable $(T/T_c)$ only, and it is shown in Fig. 1 below $T_c$.
The expansion (A36) clearly shows that near to $T_c$ the number of effective gluonic degrees of freedom
and their magnitudes will be drastically increased, though the gluon pressure initially contains only
five different massive and massless gluonic excitations.

\section{High-temperature expansion}

In order to evaluate a high-temperature expansion for the gluon pressure (A1), it is convenient to present the NP pressure (A2) as follows:

\begin{equation}
P_{NP}(T) = \Delta^2 T^2  - {6 \over \pi^2} \Delta^2 P'_1 (T) + {16 \over \pi^2} T M(T),
\end{equation}
since

\begin{eqnarray}
P_1(T) = \int_{\omega_{eff}}^{\infty} \dd \omega \ \omega \ N_g(\beta, \omega)
&=&  \int_0^{\infty} \dd \omega \ \omega N_g(\beta, \omega) - \int^{\omega_{eff}}_0 \dd \omega \ \omega
\ N_g(\beta, \omega) \nonumber\\
&=& {\pi^2 \over 6} T^2 - P'_1 (T),
\end{eqnarray}
where

\begin{equation}
\int^{\infty}_0 \dd \omega \ \omega \ N_g(\beta, \omega) = \int^{\infty}_0 \dd \omega {\omega \over e^{\beta\omega} -1} = {\pi^2 \over 6} T^2,
\end{equation}

\begin{equation}
P'_1(T) = \int^{\omega_{eff}}_0 \dd \omega \ \omega \ N_g(\beta, \omega) = \int^{\omega_{eff}}_0 \dd \omega {\omega \over e^{\beta\omega} -1},
\end{equation}
and the composition $M(T)$ is already given by the relations (A4) and (A5).

In the high-temperature limit $T \rightarrow  \infty \ (\beta = T^{-1} \rightarrow 0)$, the gluon mean number
$N_g(\beta, \omega)$ in the integral (B4) can be approximated by the corresponding series in powers of $(\beta\omega)$, since the variable $\omega$ is restricted, i.e.,

\begin{equation}
N_g(\beta, \omega) = { 1 \over e^{\beta\omega} -1} = (\beta\omega)^{-1} [ 1 - {1 \over 2} (\beta\omega) + O(\beta^2)],
\quad \beta \rightarrow 0.
\end{equation}

Let us also note in advance that in what follows for our purpose it is sufficient to keep only the positive
powers of $T$ in the evaluation of the high-temperature expansion for the gluon pressure (A1), and hence for
each term in Eq.~(A1). Thus, for the asymptotic of the integral $P'_1(T)$ up to the leading order in powers
of $T$, one obtains

\begin{equation}
P'_1(T) = \int^{\omega_{eff}}_0 \dd \omega {\omega \over e^{\beta\omega} -1} \sim T \omega_{eff},
\quad  T \rightarrow \infty.
\end{equation}

In order to investigate the behavior of the composition $M(T)$ (A4) at high temperature, it is convenient  to decompose its integral $P_2(T)$, shown in Eqs.~(A5), as follows:

\begin{equation}
P_2(T) = P_2^{(1)}(T) - P_2^{(2)}(T),
\end{equation}
where

\begin{eqnarray}
P_2^{(1)}(T) &=& \int_0^{\infty} \dd \omega \ \omega^2 \ln \left( 1- e^{-\beta\omega} \right) = - {\pi^4 \over 45} T^3 = - {\pi^2  \over 8 T} P_{SB}(T), \nonumber\\
P_2^{(2)}(T) &=& \int^{\omega_{eff}}_0 \dd \omega \ \omega^2 \ln \left( 1- e^{-\beta\omega} \right),
\end{eqnarray}
due to the relation (A30). Let us note in advance that we will not need the high-temperature asymptotic of the
integral $P_2^{(2)}(T)$.

The integral $P_3(T)$ up to the leading order in powers of $\beta \rightarrow 0$ becomes

\begin{equation}
P_3(T) = \int_0^{\omega_{eff}} \dd \omega \ \omega^2 \ln
\left( 1 - e^{- \beta\omega'} \right) \sim \int_0^{\omega_{eff}} \dd \omega \ \omega^2 \ln \beta\omega',
\quad \beta \rightarrow 0,
\end{equation}
in accordance with the expansion (B5), since the variable $\omega$ is restricted, and hence the variable
$\omega' = \sqrt{\omega^2 + a^2}$ as well, where $a= \sqrt{3} \Delta$. The last integral can be exactly calculated and the high-temperature expansion for $P_3(T)$ becomes

\begin{equation}
P_3(T) \sim {1 \over 6} \omega^3_{eff}
\ln \left( { \omega'^2_{eff} \over T^2} \right) - { 1 \over 9} \omega^3_{eff} + \Delta^2 \omega_{eff} - \sqrt{3} \Delta^3 \arctan \left({ \omega_{eff} \over \sqrt{3} \Delta} \right), \ T \rightarrow \infty,
\end{equation}
where $\omega'^2_{eff}$ is given in Eq.~(A12).

The integral $P_4(T)$ is convenient to decompose as the sum of two terms, namely

\begin{equation}
P_4(T) = P_4^{(1)}(T) + P_4^{(2)}(T),
\end{equation}
where

\begin{eqnarray}
P^{(1)}_4(T) &=& \int_{\omega_eff}^{\infty} \dd \omega \ \omega^2 \ln \left( 1 -
e^{- \beta \bar \omega} \right), \nonumber\\
P^{(2)}_4(T) &=& \int_0^{\omega_eff} \dd \omega \ \omega^2 \ln \left( 1 -
e^{- \beta \bar \omega} \right).
\end{eqnarray}

Let us begin with the integral $P^{(1)}_4(T)$, which can be re-written as follows:

\begin{equation}
P^{(1)}_4(T) = \int_{\omega_eff}^{\infty} \dd \omega \ \omega^2 \ln \left( 1 -
e^{- \beta\omega \sqrt{(1 + (a^2 / 4 \omega^2)}} \right),
\end{equation}
on account of the relation (3.6), namely $\bar \omega = \sqrt{\omega^2 + (3/4) \Delta^2} = \sqrt{\omega^2 + (a/2)^2}$. Since the variable $\omega$ is always big, then $x =(a^2 / 4 \omega^2) \ll 1$, and thus we can expand

\begin{equation}
\sqrt{(1 + x)} = 1 + {1 \over 2}x + O(x^2), \quad x \rightarrow 0.
\end{equation}
Then the integral (B13) up to the leading order in powers of small $\beta$ becomes

\begin{equation}
P^{(1)}_4(T) \sim \int_{\omega_eff}^{\infty} \dd \omega \ \omega^2 \ln \left( 1 -
e^{- \beta\omega} e^{ - (x \beta\omega / 2)} \right), \quad \beta \rightarrow 0,
\end{equation}
where the argument of the exponent $(x \beta\omega / 2) = ( a^2 / 8 \omega T) = z \ll 1$ in the
$T, \omega \rightarrow \infty$ limit, so the integral (B15) can be present as follows:

\begin{equation}
P^{(1)}_4(T) \sim \int_{\omega_eff}^{\infty} \dd \omega \ \omega^2 \ln \left( 1 -
e^{- \beta\omega} [ 1 - z + O(z^2)]  \right), \quad z \ll 1, \quad \beta \rightarrow 0,
\end{equation}
or, equivalently,

\begin{eqnarray}
P^{(1)}_4(T) &\sim& \int_{\omega_eff}^{\infty} \dd \omega \ \omega^2 \ln \left[ \left( 1 -
e^{- \beta\omega} \right) \left(  1 + { z  \over e^{\beta\omega} -1} \right)  \right] \nonumber\\
&\sim& P_2(T) + \int_{\omega_eff}^{\infty} \dd \omega \ \omega^2 \ln \left(  1 + { z  \over e^{\beta\omega} -1} \right) \sim P_2(T) + P'_2(T),
\end{eqnarray}
as it follows from equations (A5) or (B7)-(B8).
The argument of the logarithm in the second integral is again always small $(z / e^{\beta\omega} -1) \ll 1$
in the $T, \omega \rightarrow \infty$ limit, and thus we can expand it and obtain in the leading order

\begin{equation}
P'_2(T) = \int_{\omega_eff}^{\infty} \dd \omega \ \omega^2 \ln \left(  1 + { z  \over e^{\beta\omega} -1} \right)
\sim  { a^2 \over 8} \beta \int_{\omega_eff}^{\infty} \dd \omega \ { \omega \over e^{\beta\omega} -1}.
\end{equation}
The last integral is nothing but $P_1(T)$ defined in Eq.~(B2), so that combining
(B2) and (B6), one obtains

\begin{equation}
P'_2(T) \sim { a^2 \over 8} \beta P_1(T)
\sim  {\pi^2 \over 16} \Delta^2 T - {3 \over 8} \Delta^2 \omega_{eff},
\quad T \rightarrow \infty,
\end{equation}
and then the high-temperature expansion for $P^{(1)}_4(T)$ becomes

\begin{equation}
P^{(1)}_4(T) \sim P_2(T) +  {\pi^2 \over 16} \Delta^2 T - {3 \over 8} \Delta^2 \omega_{eff},
\quad T \rightarrow \infty.
\end{equation}

The integral $P_4^{(2)}(T)$ in the leading order in powers of $\beta \rightarrow 0$ becomes

\begin{equation}
P_4^{(2)}(T) = \int_0^{\omega_{eff}} \dd \omega \ \omega^2 \ln \left( 1 -
e^{- \beta \bar \omega} \right) \sim \int_0^{\omega_{eff}} \dd \omega \ \omega^2 \ln \beta \bar \omega,
\quad \beta \rightarrow 0,
\end{equation}
in accordance with the expansion (B5), since the variable $\omega$ is restricted, and hence the variable
$\bar \omega = \sqrt{\omega^2 + (a/2)^2}$ as well, where again $a= \sqrt{3} \Delta$.
From the relations (3.4) and (3.5) it follows that $\omega' \rightarrow \bar \omega$ by $\Delta \rightarrow (1/2) \Delta$, so making this replacement in the expansion (B10), one automatically obtains the high-temperature
 expansion for the integral $P_4^{(2)}(T)$ as follows:

\begin{eqnarray}
P_4^{(2)} \sim {1 \over 6} \omega^3_{eff}
\ln \left( { \bar \omega^2_{eff} \over T^2} \right) - { 1 \over 9} \omega^3_{eff} + { 1 \over 4} \Delta^2 \omega_{eff} - { \sqrt{3} \over 8} \Delta^3 \arctan \left({ 2 \omega_{eff} \over \sqrt{3} \Delta} \right),
T \rightarrow \infty,
\end{eqnarray}
where $\bar \omega^2_{eff} = \omega^2_{eff} + (3 /4) \Delta^2$.

The high-temperature expansion for the composition (A4), on account of the relations (B11)-(B12) and the
previous expansions (B10), (B20) and (B22) and after doing some algebra, becomes

\begin{eqnarray}
{16 \over \pi^2} T M(T) &\sim& {18 \over \pi^2} \Delta^2 \omega_{eff} T - \Delta^2 T^2
+ {8 \over 3 \pi^2} \omega^3_{eff} T
\ln \left( { \omega'^2_{eff} \over \bar \omega^2_{eff}} \right)
\nonumber\\
&+& {2 \sqrt{3} \over \pi^2} \Delta^3 T\arctan \left({ 2 \omega_{eff} \over \sqrt{3} \Delta} \right) - {16 \sqrt{3} \over \pi^2} \Delta^3 T \arctan \left({ \omega_{eff} \over \sqrt{3} \Delta} \right), \ T \rightarrow \infty.
\nonumber\\
\end{eqnarray}
Substituting this expansion into the Eq.~(B1), and on account of the expansion (B6), we obtain

\begin{eqnarray}
P_{NP}(T) &\sim& {12 \over \pi^2} \Delta^2 \omega_{eff}T  + {8 \over 3 \pi^2} \omega^3_{eff} T
\ln \left( { \omega'^2_{eff} \over \bar \omega^2_{eff}} \right) \nonumber\\
&+& {2 \sqrt{3} \over \pi^2} \Delta^3 T\arctan \left({ 2 \omega_{eff} \over \sqrt{3} \Delta} \right) - {16 \sqrt{3} \over \pi^2} \Delta^3 T \arctan \left({ \omega_{eff} \over \sqrt{3} \Delta} \right), \
T \rightarrow \infty. \nonumber\\
\end{eqnarray}
One concludes that the exact cancelation of the $\Delta^2 T^2$ term occurs within the NP pressure itself. Thus $P_{NP}(T) \sim T$ up to the leading order in the $T \rightarrow \infty$ limit. At the same time, at high temperatures the exact cancelation of the $P_{SB}(T)$ term occurs within the composition $(16 / \pi^2) T M_1(T)= (16 / \pi^2) T [ P_2(T)-P_4(T)]$, which enters the composition (B.23).
To show this explicitly, let us substitute into the former composition the relation (B7), on account
of the relations (B8), and the relation (B11), on account of the expansion (B20), and doing some algebra, one obtains

\begin{eqnarray}
{16 \over \pi^2} T M_1(T) &\sim& -2P_{SB}(T) + 2 P_{SB}(T) - \Delta^2 T^2 + {6 \over \pi^2} \Delta^2 \omega_{eff} T
- {16 \over \pi^2} T P_4^{(2)}(T) \nonumber\\
&\sim& - \Delta^2 T^2 + {6 \over \pi^2} \Delta^2 \omega_{eff} T - {16 \over \pi^2} T P_4^{(2)}(T),
\quad T \rightarrow \infty,
\end{eqnarray}
from which the above-mentioned exact cancelation explicitly follows. The exact cancelation of the
$P_2^{(2)}(T)$ terms and the expansion (B22) for the  $P_4^{(2)}(T)$ term are not shown, for simplicity.

Let us now consider Eq.~(6.4), which is convenient to decompose as follows:

\begin{equation}
P^s_{PT}(T) = {9 \over 2 \pi^2} \alpha_s \Delta^2
\int_{\Lambda_{YM}}^{\infty} \dd\omega \ \omega^2 \
{ 1 \over \bar \omega} { 1 \over e^{\beta \bar \omega}  - 1} = P^s_1(T) -  P^s_2(T),
\end{equation}
where

\begin{eqnarray}
P^s_1(T) &=& {9 \over 2 \pi^2} \alpha_s \Delta^2
\int_0^{\infty} \dd\omega \ \omega^2 \
{ 1 \over \bar \omega} { 1 \over e^{\beta \bar \omega}  - 1}, \nonumber\\
P^s_2(T) &=& {9 \over 2 \pi^2} \alpha_s \Delta^2
\int^{\Lambda_{YM}}_0 \dd\omega \ \omega^2 \
{ 1 \over \bar \omega} { 1 \over e^{\beta \bar \omega}  - 1},
\end{eqnarray}
and let us remind that
$\bar \omega = \sqrt{\omega^2 + (3/4) \Delta^2} = \sqrt{\omega^2 + (a/2)^2}$.

In the integral $P^s_1(T)$ it is convenient to introduce a new dimensionless variable
$x = \beta \bar \omega = \beta \sqrt{\omega^2 + (a/2)^2}$. After doing some algebra, it becomes

\begin{equation}
P^s_1(T) = {9 \over 2 \pi^2} \alpha_s \Delta^2
\int_0^{\infty} \dd\omega \ \omega^2 \
{ 1 \over \bar \omega} { 1 \over e^{\beta \bar \omega}  - 1} =
{9 \over 2 \pi^2} \alpha_s \Delta^2 T^2
\int_{(\beta a/2)}^{\infty} \dd x \ { \sqrt{ x^2 - (\beta a/2)^2} \over e^x  - 1}.
\end{equation}
The last integral when $\beta \rightarrow 0$ can be approximated up to the leading order as follows:

\begin{equation}
\int_{(\beta a/2)}^{\infty} \dd x \ { \sqrt{ x^2 - (\beta a/2)^2} \over e^x  - 1} \sim
\int_0^{\infty} \dd x \ { x \over e^x  - 1} = {\pi^2 \over 6}, \quad \beta \rightarrow 0,
\end{equation}
then for the integral $P^s_1(T)$ up to the leading order in powers of $T$ we get

\begin{equation}
P^s_1(T) \sim {3 \over 4} \alpha_s \Delta^2 T^2, \quad T \rightarrow \infty.
\end{equation}

In the integral $P^s_2(T)$ the variable $\omega$ is restricted, and hence $\bar \omega$ as well.
So up to the leading order in the $ T \rightarrow \infty \ (\beta = T^{-1} \rightarrow 0)$ limit this integral can be approximated as

\begin{eqnarray}
P^s_2(T) = {9 \over 2 \pi^2} \alpha_s \Delta^2
\int^{\Lambda_{YM}}_0 \dd\omega \ \omega^2 \
{ 1 \over \bar \omega} { 1 \over e^{\beta \bar \omega}  - 1} \sim {9 \over 2 \pi^2} \alpha_s \Delta^2 T
\int^{\Lambda_{YM}}_0 \dd\omega \ { \omega^2 \over \bar \omega^2}, \quad T \rightarrow \infty, \nonumber\\
\end{eqnarray}
in accordance with the expansion (B5).
The last integral can easily be integrated and thus the high-temperature expansion for the
$P^s_2(T)$ term looks like

\begin{equation}
P^s_2(T) \sim {9 \over 2 \pi^2} \alpha_s \Delta^2 T \left[ \Lambda_{YM} - {\sqrt{3} \over 2} \Delta
\arctan \left( {2 \Lambda_{YM} \over \sqrt{3} \Delta} \right) \right], \quad T \rightarrow \infty.
\end{equation}

Summing it with the expansion (B30), for the integral (B26) one obtains

\begin{equation}
P^s_{PT}(T) \sim {9 \over 2 \pi^2} \alpha_s \Delta^2 \left[ {\pi^2 \over 6} T^2 -
T \left( \Lambda_{YM} - {\sqrt{3} \over 2} \Delta
\arctan \left( {2 \Lambda_{YM} \over \sqrt{3} \Delta} \right) \right) \right], \quad T \rightarrow \infty,
\end{equation}
which is nothing but the high-temperature expansion for the $\alpha_s$-dependent PT part of the gluon pressure.

The high-temperature expansion of the gluon pressure is to be obtained by summing up the expansions (B24)
and (B33), so it is

\begin{eqnarray}
P_g(T) &=& [P_{NP}(T) + P^s_{PT}(T)] \sim {12 \over \pi^2} \Delta^2 \omega_{eff}T  + {8 \over 3 \pi^2} \omega^3_{eff} T \ln \left( { \omega'_{eff}\over \bar \omega_{eff} } \right)^2 \nonumber\\
&+& {2 \sqrt{3} \over \pi^2} \Delta^3 T\arctan \left({ 2 \omega_{eff} \over \sqrt{3} \Delta} \right) - {16 \sqrt{3} \over \pi^2} \Delta^3 T \arctan \left({ \omega_{eff} \over \sqrt{3} \Delta} \right) \nonumber\\
&+& {9 \over 2 \pi^2} \alpha_s \Delta^2 \left[ {\pi^2 \over 6} T^2 -
T \left( \Lambda_{YM} - {\sqrt{3} \over 2} \Delta \arctan \left( {2 \Lambda_{YM} \over \sqrt{3} \Delta} \right) \right) \right], \ T \rightarrow \infty. \nonumber\\
\end{eqnarray}
Let us emphasize that the high-temperature expansions for all three pressures (B24), (B33) and (B34) non-analytically depend on the mass gap in terms $\sim \Delta^3 T \sim (\Delta^2)^{(3/2)}T$, but it is not an expansion parameter like $\alpha_s$. From asymptotics (B33) and (B34) it follows that $P^s_{PT}(T)$, and hence $P_g(T)$,
behaves like $T^2$ in the leading order, while remembering $P_{NP}(T) \sim T$, see expansion (B24).
It is also interesting to note that the effective massive gluonic "excitations"
$\omega'_{eff} = \sqrt{\omega^2_{eff} + 3 \Delta^2}$ and $\bar \omega_{eff} = \sqrt{\omega^2_{eff} + (3/4) \Delta^2}$ are logarithmically suppressed at high temperatures, while there is no dependence on the effective massive gluonic "excitation"  $\tilde{\omega}_{eff} = \sqrt{\Lambda^2_{YM} + (3/4) \Delta^2}$ at all.

In a more compact form the previous expansion looks like

\begin{equation}
P_g(T) \sim B_2 \alpha_s \Delta^2 T^2 + [B_3 \Delta^3 + M^3]T, \quad T \rightarrow \infty,
\end{equation}
where $M^3$ denotes the terms of the dimensions of the $\GeV^3$, which depend analytically on the
mass gap $\Delta^2$. The explicit expressions for it and for both constants $B_2$ and $B_3$ can be easily
restored from the expansion (B34), if necessary.

\end{document}